\documentclass[aps,pra,twocolumn,superscriptaddress, showpacs,halfspacing,10pt]{revtex4-1}

\usepackage[]{longtable}
\usepackage{graphicx}
\usepackage{todonotes}
\usepackage{braket}
\usepackage{amsmath}
\usepackage{upgreek}
\usepackage{here}
\usepackage[final]{changes}

\begin{document}

\newcommand{\NZIAS}{
Centre for Theoretical Chemistry and Physics,
The New Zealand Institute for Advanced Study,
Massey University Auckland, Private Bag 102904, 0745 Auckland, New Zealand}

\newcommand{\TAU}{
School of Chemistry, Tel Aviv University, 69978 Tel Aviv, Israel}

\newcommand{\HIM}{
Helmholtz Institute Mainz, Mainz D-55128, Germany}

%\title{Benchmarking \emph{ab initio} Fock-space coupled-cluster calculations by optical spectroscopy of Sn$^{11+...14+}$ ions}

%\title{Re-evaluation of the Fine Structure Splitting of Sn$^{11+...14+}$ Ions by Optical Spectroscopy}

\title{Analysis of the fine structure of Sn$^{11+...14+}$ ions by optical spectroscopy \\ in an electron beam ion trap}

\author{A.~Windberger}
\affiliation{Advanced Research Center for Nanolithography, Science Park~110, 1098~XG Amsterdam, The Netherlands}
\affiliation{Max-Planck-Institut f\"ur Kernphysik, Saupfercheckweg 1, 69117 Heidelberg, Germany}
\author{F.~Torretti}
\affiliation{Advanced Research Center for Nanolithography, Science Park~110, 1098~XG Amsterdam, The Netherlands}
\affiliation{Department of Physics and Astronomy, and LaserLaB, Vrije Universiteit, De Boelelaan 1081, 1081 HV Amsterdam, The Netherlands}
\author{A.~Borschevsky}
\affiliation{The Van Swinderen Institute, University of Groningen, Nijenborgh 4, 9747 AG Groningen, The Netherlands}
%\affiliation{\HIM}
\author{A.~Ryabtsev}
\affiliation{Institute of Spectroscopy, Russian Academy of Sciences, Troitsk, Moscow, 108840 Russia}
\affiliation{EUV Labs, Ltd., Troitsk, Moscow, 108840 Russia}
\author{S.~Dobrodey}
\affiliation{Max-Planck-Institut f\"ur Kernphysik, Saupfercheckweg 1, 69117 Heidelberg, Germany}
\author{H.~Bekker}
\affiliation{Max-Planck-Institut f\"ur Kernphysik, Saupfercheckweg 1, 69117 Heidelberg, Germany}
\author{E.~Eliav}
\affiliation{\TAU}
\author{U.~Kaldor}
\affiliation{\TAU}
\author{W.~Ubachs}
\affiliation{Advanced Research Center for Nanolithography, Science Park~110, 1098~XG Amsterdam, The Netherlands}
\affiliation{Department of Physics and Astronomy, and LaserLaB, Vrije Universiteit, De Boelelaan 1081, 1081 HV Amsterdam, The Netherlands}
\author{R.~Hoekstra}
\affiliation{Advanced Research Center for Nanolithography, Science Park~110, 1098~XG Amsterdam, The Netherlands}
\affiliation{Zernike Institute for Advanced Materials, University of Groningen, Nijenborgh 4, 9747 AG Groningen, The Netherlands}
\author{J.~R.~Crespo~L\'{o}pez-Urrutia}
\affiliation{Max-Planck-Institut f\"ur Kernphysik, Saupfercheckweg 1, 69117 Heidelberg, Germany}
\author{O.~O.~Versolato}
\affiliation{Advanced Research Center for Nanolithography, Science Park~110, 1098~XG Amsterdam, The Netherlands}
\email{o.versolato@arcnl.nl}


\date{\today}
\normalem
\begin{abstract}
We experimentally re-evaluate the fine structure of Sn$^{11+...14+}$ ions. These ions are essential in bright extreme-ultraviolet (EUV) plasma-light sources for next-generation nanolithography, but their complex electronic structure is an open challenge for both theory and experiment. We combine optical spectroscopy of magnetic dipole $M1$ transitions, in a wavelength range covering 260\,nm to 780\,nm, with charge-state selective ionization in an electron beam ion trap. Our measurements confirm the predictive power of \emph{ab initio} calculations based on Fock space coupled cluster theory. We validate our line identification using semi-empirical Cowan calculations with adjustable wavefunction parameters. Available Ritz combinations further strengthen our analysis. Comparison with previous work suggests that line identifications in the EUV need to be revisited.
\end{abstract}

%\keywords{}

\maketitle

\section{Introduction}
The strongly correlated electronic structure of heavy, multi-electron open shell ions is notoriously difficult to calculate and their complicated structure furthermore hampers straightforward experimental assessment.
A typical example of this class of systems are Sn ions in charge states 7+ through 14+ with their open [Kr]$4d$ shell structure. These specific ions are used to generate extreme ultraviolet (EUV) light at 13.5\,nm wavelength in laser-produced-plasma (LPP) sources for nanolithographic applications \mbox{\cite{Benschop2008,Banine2011}}. The EUV light is generated by thousands of transitions that form so-called unresolved transition arrays (UTAs) with little dependence on the charge state of the ion. For the relevant [Kr]$4d^m$ tin ions, with $m$=6-0\,\cite{OSullivan2015}, the EUV-contributing upper configurations are 4$p^6$4$d^{m-1}$4$f^1$, 4$p^6$4$d^{m-1}$5$p^1$, and 4$p^5$4$d^{m+1}$. The sheer multitude of lines in these UTAs complicates their accurate identification. Spectroscopic work using discharge sources \cite{Azarov1993,tolstikhina2006ATOMICDATA,churilov2006SnVIII,churilov2006SnXIII--XV,ryabtsev2008SnXIV,churilov2006SnIX--SnXII}, laser-produced-plasmas \cite{Svendsen1994}, or tokamaks \cite{sugar1991resonance,sugar1992rb} is challenging as the UTAs of various Sn ions are strongly blended. Nevertheless, work on discharge sources provides the most accurate spectroscopic data to date for highly charged Sn ions.

Besides experiments on thermal plasmas, there is work on charge state-resolved spectroscopy in the EUV regime. Charge-exchange spectroscopy (CXS) was performed by means of Sn ion beams colliding on He\,\cite{ohashi2009complementary,DArcy2009transitions,DArcy2009identification,ohashi2010euv} and Xe\,\cite{ohashi2010euv} gas targets. The spectral accuracy achieved in those studies was lower than that of the discharge work due to instrument resolution. Studies using electron beam ion traps (EBITs)\,\cite{ohashi2009complementary,yatsurugi2011euv} which also provide charge state selectivity were similarly limited. In addition, the EBIT studies focused on the higher charge states of tin, outside of the range most relevant for EUV plasma sources. For these reasons, none of the charge-state-resolved studies could so far directly challenge the spectral data obtained from discharge sources.

\begin{figure}[t]
\includegraphics{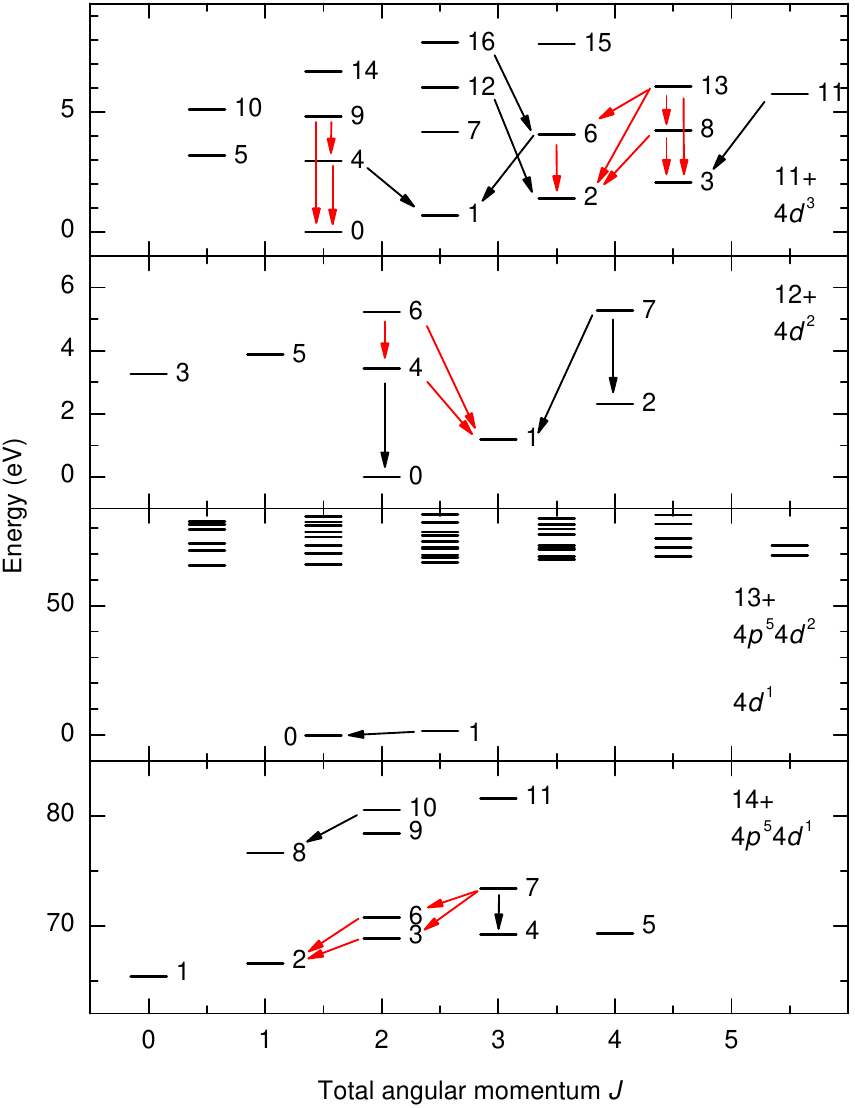}
\caption{Grotrian diagrams obtained from Flexible Atomic Code calculations \cite{gu2008flexible} depicting the atomic structure of the lowest energy configurations exhibiting optical magnetic dipole $M1$ transitions of the Sn charge states investigated in this work. The closed-shell [Kr] ground state of Sn$^{14+}$ has no splitting; instead the first excited configuration is shown. The levels are numbered following their energy ordering. Their respective term symbols can be found in Table\,\ref{all_data}. Identified transitions are marked by arrows: (red) transitions confirmed by Ritz combinations; (black) the remainder. All observed spectral lines and identified transitions are listed in Table\,\ref{tab:lines}.
\label{fig:sn_level_structure}}
\end{figure}

The above issues of unresolved transitions in UTAs and limited resolution can be circumvented by turning to the optical range and addressing the optical magnetic dipole ($M$1) transitions in between fine structure levels in the ground electronic configuration (see Fig.\,\ref{fig:sn_level_structure}). This approach eliminates the uncertainties introduced by the reconstruction of ground state levels using a Ritz procedure based on the measured EUV lines. Such optical lines of Sn in charge states beyond 3+ have not been identified before.

We present results from charge state-resolved optical spectroscopy on Sn$^{11+...14+}$ using FLASH-EBIT \cite{epp2010x} at the Max Planck Institute for Nuclear Physics in Heidelberg. Tuning the energy of the electron beam enables us to assign each of the observed spectral lines to specific charge states. The $M$1 transitions are assigned using \textit{ab initio} Fock space coupled cluster (FSCC) calculations and we confirm them with the help of the semi-empirical Cowan code \cite{cowan1981}, which allows to adjust wave function scaling parameters to fit the observed spectra. 

Optical transitions in heavy multi-electron open shell ions like Sn$^{11+...14+}$ represent a stringent test for \textit{ab initio} atomic structure calculations of strongly correlated many-electron ions. These transitions have non-negligible Breit contributions. Application of FSCC theory to this type of problems where multiple vacancies with high angular momenta can couple to each other is a promising approach. Coupled cluster methods have found wide application since the 1950s to analogous problems found in nuclear physics and quantum chemistry, yet their use for atomic physics problems involving highly charged ions is more recent (see, e.g., Ref.\,\cite{lindgren1996relativistic}). There are only few such works so far although the methods consistently show good agreement with experiment (see, e.g., Refs.\,\cite{eliav1995relativistic,nandy2013development,safronova2014highly,windberger2015}). It is therefore instructive to explore their \emph{ab initio} performance in specific cases where other established calculational tools require empirical adjustments and judicious choices of configurations sets in order to analyze spectral data.
Therefore, with optical spectroscopic studies we can address the issue of a precise theory-experiment comparison in complex open-shell systems without the high spectral density that has to be faced in EUV spectroscopy work. In the present case, both the important practical applications of the ions under study as well as the relative novelty of using FSCC calculations for them make such a comparison particularly valuable. 

Our data enable a re-evaluation of the fine structure splitting of Sn$^{11+...14+}$ ions and provide a benchmark for state-of-the-art atomic structure calculations such as FSCC. We infer that the identification of EUV lines needs to be revisited in previous works such as Ref.\,\cite{churilov2006SnIX--SnXII,churilov2006SnXIII--XV} as was also suggested previously in Ref.\,\cite{ryabtsev2011resonance}. The line identifications in those works constitute the basis for plasma modeling of EUV light sources.
%
%On the theory side, over the decades and with a tremendous effort, a great variety of theoretical methods have been applied to them, and various refinements have been implemented in those. Still, the \emph{ab initio} predictive power of electronic structure calculations is limited, and uncertainties in calculated energies in the order of several percent are common. This means, e.g., that the complicated, dense photon spectra which arise when semi-filled \emph{d} or \emph{f} subshells are present still defy not only the experimental techniques in the laboratory, but the theoretical analysis as well.

\section{Experiment} 
We produced and trapped Sn ions with FLASH-EBIT\,\cite{epp2010x} using a mono-energetic electron beam to bring them to the desired charge state. The beam energy was controlled by the acceleration potential applied between the emitting cathode and the central trap drift tube. High current densities were reached by compressing the electron beam down to a diameter of approximately 50\,$\upmu$m using the 6\,T magnetic field produced by a pair of superconducting Helmholtz-coils. 
A tenuous, well collimated molecular beam generated by the evaporation of volatile tetra-i-propyltin (C$_{12}$H$_{28}$Sn) was the carrier of Sn atoms into the electron beam, which dissociated the molecule and preferentially trapped the heavy Sn ions thereby produced.	While the lighter atoms and ions of C and H escaped from the trap, Sn ions remained trapped, radially by the space charge potential of the electron beam and axially by a potential generated by a set of drift tubes. 

The trapped ions were electronically excited by electron impact to a manifold of states, many of them close to the ionization continuum since the beam energy is close to the respective ionization threshold.
%
%\begin{figure}[t]
%\includegraphics[width=8.66cm]{images/setup_v1.pdf}
%\caption{Schematic of the optical spectrometer setup. Two lenses relay a 1:1 image of the ion cloud outside of the vacuum chamber of the EBIT. A periscope box rotates the horizontal ion cloud image by 90$^{\circ}$ and focuses it onto the vertical entrance slit of a spectrometer. A dispersed image of the spectrometer entrance slit is acquired by a CCD chip.
%\label{fig:spectrometer}}
%\end{figure}%
%
Subsequent fluorescent cascades down towards the ground state cover a broad spectral range. This light was focused onto the entrance slit of a 320\,mm-focal-length spectrometer employing a 300\,lines/mm grating. We use in this work a low groove density grating and a short focal distance instrument for the convenience of having a broad spectral coverage, given the large number of spectra to be acquired. The spectral image recorded by a cooled CCD is integrated along the nondispersive axis after correcting for optical aberrations and removing cosmic muon events. Line widths of typically $\sim$1\,nm (full-width-at-half-maximum) were obtained near 400\,nm wavelength. For calibration a Hg or a Ne-Ar lamp was placed in front of the spectrometer entrance slit.  

A typical acquisition cycle consisted of a short calibration and a series of 30\,min exposures. After each acquisition the electron beam acceleration potential was increased by 10\,V starting from a minimum acceleration potential of 137\,V and ending at 477\,V. The electron beam current was kept at a constant 10\,mA. The dense electron beam produces a strong space charge potential which is partially compensated by the trapped ions \cite{penetrante1991evolution,brenner2007lifetime}. This reduces the acceleration potential by $\sim$20-40\,V \cite{Bekker2015a} to yield the actual electron beam energy in the interaction region. The chosen range of the acceleration potential enabled the production of charge states from Sn$^{7+}$, with its ionization potential (IP) of 135\,V \cite{rodrigues2004systematic,NIST_ASD} up to Sn$^{16+}$ with IP=437\,V. We focus on the charge states Sn$^{11+...14+}$ that could be reliably identified.

After stepping the EBIT acceleration potential through the full voltage range, the rotatable grating was set to cover an 125\,nm adjacent wavelength range. Typically, a range of 270\,nm was recorded at each grating position such that the different regions overlapped. Next, the acceleration potential was stepped through again. This procedure was repeated for three settings of the spectrometer grating to cover the full accessible wavelength range from 260\,nm to 780\,nm.

\begin{figure*}[]
\includegraphics[width=\textwidth]{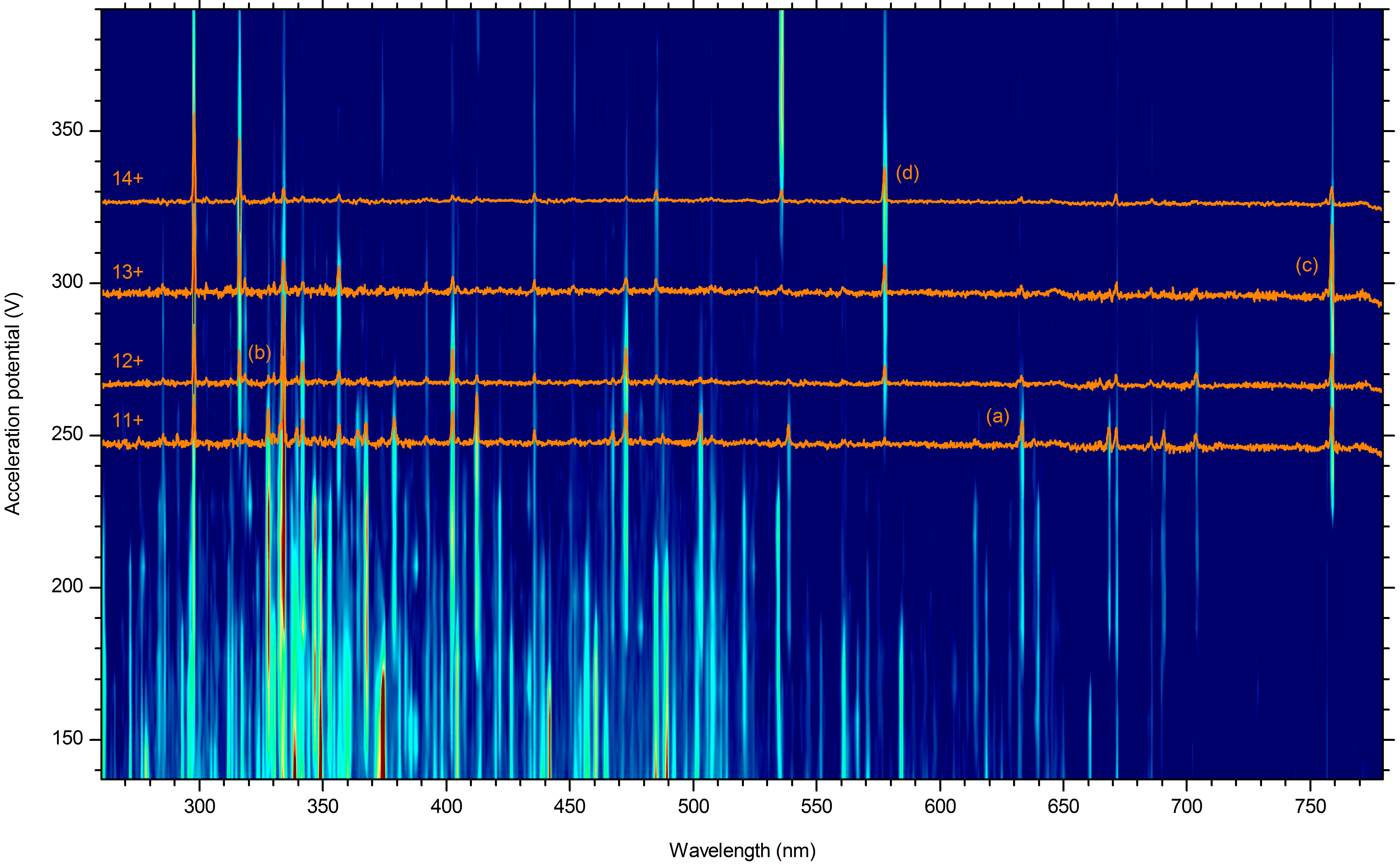}
\caption{Composite spectral map of Sn ions interpolated between discrete voltage steps of the EBIT acceleration potential at a 10\,mA beam current. The orange curves represent spectra of Sn$^{11+...14+}$ ions taken at a maximum of the fluorescence yield of a specific charge state; they are individually scaled for better visibility. Enlarged sections around the lines indicated by symbols (a), (b), (c), and (d) can be found in Fig.\,\ref{fig:optical_zooms} to visualize the identification of charge states through their fluorescence profiles.
\label{fig:Sn_map_charge_states}}
\end{figure*}%

\section{Theory}
Two calculational methods are compared in this work. First, we present dedicated \emph{ab initio} FSCC calculations and show the accuracy of these predictions by comparison with our experimental data. Second, we compare them with Cowan-code calculations using empirically adjusted wavefunction scaling parameters. This code is a mature tool used to identify lines, and serves particularly well when a combination of experimental observations provides additional data on electronic energy levels for their empirical adjustment. We also use it to obtain weighted transition rates $gA$ used to predict line strengths in the following. When necessary, auxiliary calculations were carried out with the Flexible Atomic Code (FAC)\,\cite{gu2008flexible}. 

\subsection{Fock space coupled cluster}
Calculations of the transition energies were performed for the ions of interest using the FSCC method within the framework of the projected Dirac-Coulomb-Breit Hamiltonian~\cite{Suc80},
\begin{eqnarray}
H_{DCB}= \displaystyle\sum\limits_{i}h_{D}(i)+\displaystyle\sum\limits_{i<j}(1/r_{ij}+B_{ij}).
\label{eqHdcb}
\end{eqnarray}
Here, $h_D$ is the one-electron Dirac Hamiltonian,
\begin{eqnarray}
h_{D}(i)=c\pmb{\alpha }_{i}\cdot \mathbf{p}_{i}+c^{2}\beta _{i}+V_{nuc}(i),
\label{eqHd}
\end{eqnarray}
where $\pmb{\alpha}$ and $\beta$ are the four-dimensional Dirac matrices. The nuclear potential $V_{nuc}(i)$  takes into account the finite size of the nucleus, modeled by a uniformly charged sphere \cite{IshBarBin85}. The two-electron term includes the nonrelativistic electron repulsion and the frequency independent Breit operator,
\begin{eqnarray}
B_{ij}=-\frac{1}{2r_{ij}}[\pmb{\alpha }_{i}\cdot \pmb{\alpha }_{j}+(%
\pmb{\alpha }_{i}\cdot \mathbf{r}_{ij})(\pmb{\alpha }_{j}\cdot \mathbf{%
r}_{ij})/r_{ij}^{2}],
\label{eqBij}
\end{eqnarray}
and is correct to second order in the fine structure constant. The calculations for Sn$^{14+}$, Sn$^{13+}$, and Sn$^{12+}$ started from the closed shell reference 4$s^{2}$4$p^6$ configuration of Sn$^{14+}$. In the current state of the code, atomic systems with a maximum of two open shell electrons/holes can be treated, which does not apply to Sn$^{11+}$ with its 4$s^{2}$4$p^6$4$d^3$ ground state configuration. After the first stage of the calculation, consisting of solving the relativistic Hartree-Fock equations and correlating the closed shell reference state, different FSCC schemes were used for the different ions. 
%Two electrons were removed from the 4p orbital to obtain the transition energies of Sn$^{16+}$. 
In case of Sn$^{14+}$, a single electron was excited from  the 4$p$ to the 4$d$ orbital to reach the 4$p^5$4$d^1$ configuration. For Sn$^{12+}$, two electrons were added to the closed shell reference state. In this calculation, to achieve optimal accuracy, a large model space was used, comprised of three $s$, three $p$, three $d$, three $f$, two $g$, and one $h$ orbitals. The intermediate Hamiltonian method was employed to facilitate convergence \cite{EliVilIsh05}. The fine structure splitting of Sn$^{13+}$ was also obtained in the framework of this calculation, as a result of adding the first electron to the closed shell reference state. 

The uncontracted universal basis set \cite{MalSilIsh93} was used for all the ions, consisting of 37 $s$, 31 $p$, 26 $d$, 21 $f$, 16 $g$, 11 $h$, and six $i$ functions; the convergence of the obtained transition energies with respect to the size of the basis set was verified. All the electrons were correlated. The FSCC calculations were performed using the Tel-Aviv Relativistic Atomic FSCC code (TRAFS-3C).

Lamb shifts for the various levels were obtained using the recently developed effective potential method, implemented in the QEDMOD program \cite{Shabaev2015_qed}. Here, an important feature is the inclusion of both the vacuum polarization and the self-energy components of the Lamb shift into the self-consistent field procedure. Thus, together with the first order QED interactions, many important higher order terms are included in the final Lamb-shift expression. Its contribution to level energies is typically 20-60\,cm$^{-1}$. The results of the calculations are presented in Table\,\ref{all_data} where they are compared to the experimental results, as will be discussed in the next section. 

\subsection{Cowan}\label{Cowan}
In the Cowan code \cite{cowan1981}, radial wavefunctions are obtained with a Hartree-Fock relativistic (HFR) method using a correlation term in the potential but neglecting Breit interaction. With these wavefunctions, the electrostatic single configuration radial integrals $F^k$ and $G^k$ (Slater integrals), configuration-interaction Coulomb radial integrals, and spin-orbit parameters $\zeta$ are then calculated. From these, the energy levels and intermediate coupling eigenvectors are extracted. Subsequently, \emph{ab initio} values are obtained for the wavelengths and transition probabilities. However, the resulting energy level splittings are generally larger than those observed because of the cumulative influence of a large number of small perturbations originating from configuration interactions. To compensate for these effects the \emph{ab initio} values of the electrostatic integrals are empirically scaled down by a factor between 0.7 and 0.95, depending on the charge state. Spin-orbit parameters can be also scaled. Both scalings are needed for a semi-empirical adjustment of the predicted spectrum that can be performed if enough experimental levels are available. The electrostatic and spin-orbit integrals are then adjusted to give the best possible fit of the calculated eigenvalues to the observed energy levels. Effective Coulomb-interaction operators $\alpha$, $\beta$, and $T$. are added as fit parameters to represent weak configuration-interaction corrections to the electrostatic single configuration effects. The results of this semi-empirical adjustment procedure are more useful for the interpretation of a particular experimental spectrum than the \emph{ab initio} Cowan calculation. Additionally, the empirical ratios of the fitted (FIT) to the HFR energy parameters can be extrapolated, e.g., to neighboring ions along an isoelectronic sequence to improve the reliability of \emph{ab initio} predictions.

\section{Results} 
\begin{figure*}
\includegraphics[width=\textwidth]{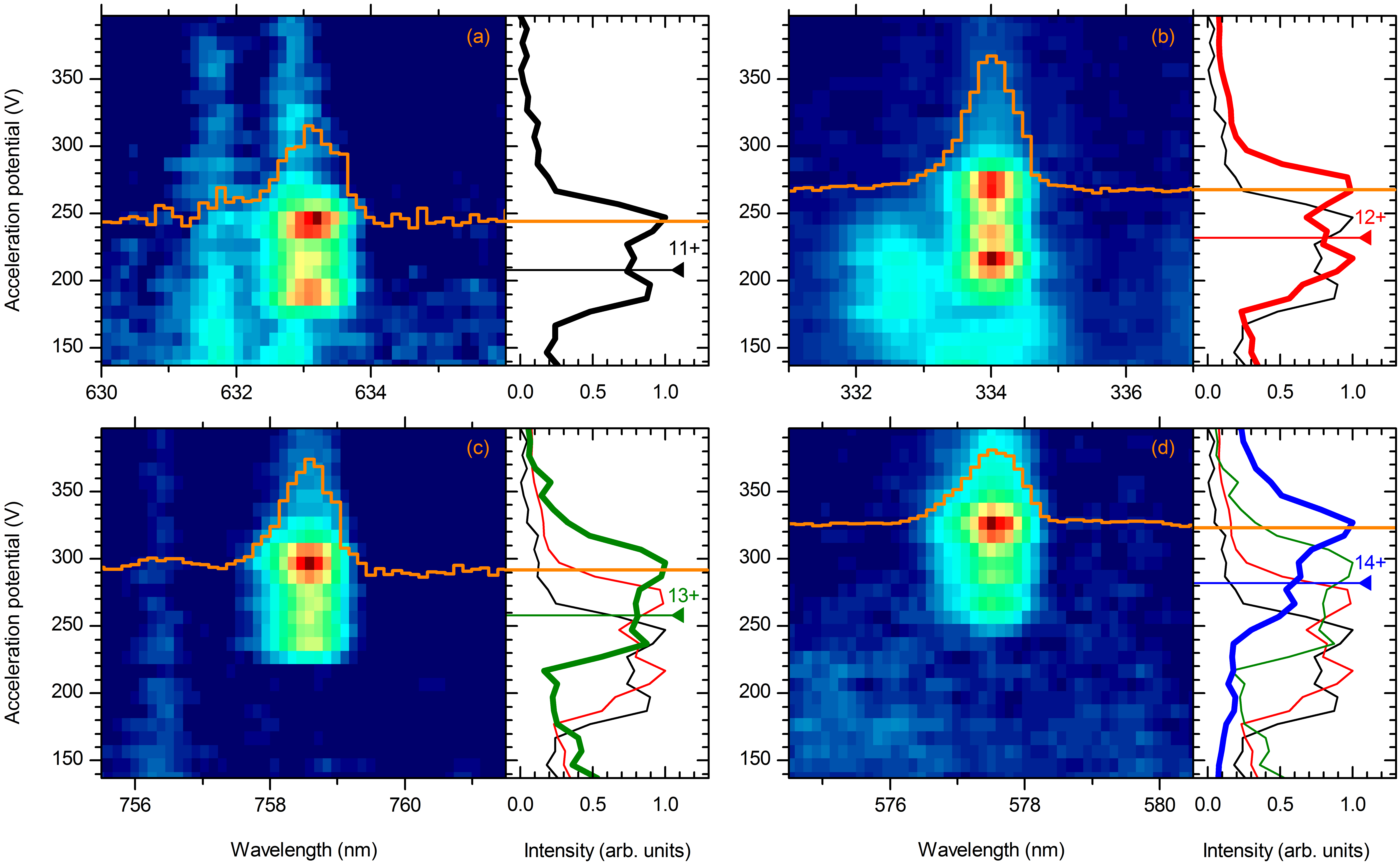}
\caption{Enlarged sections around the spectral lines marked (a), (b), (c), and (d) in Fig.\,\ref{fig:Sn_map_charge_states} to facilitate charge state identification. On the right hand side of the composite spectral maps we show the result of their projection onto the acceleration potential axis uncorrected for space charge. The profiles are averaged over several bright lines exhibiting the same voltage dependence. A charge state ``ladder'' is produced (see main text). The overlaid orange spectra (also see Fig. \ref{fig:Sn_map_charge_states}) are obtained at the acceleration potential which maximizes the fluorescence yield.  Theoretical ionization energies \cite{rodrigues2004systematic,NIST_ASD} are depicted by triangles.
\label{fig:optical_zooms}}
\end{figure*}

In the following, we present optical spectra of tin ions in charge states Sn$^{11+...14+}$ obtained in a charge-state-resolved manner (see Figs.\,\ref{fig:Sn_map_charge_states} and \ref{fig:optical_zooms}). We interpret the data using FSCC predictions and semi-empirical Cowan-code calculations. Furthermore, we perform a comparison with predicted optical transition energies inferred from existing data in the EUV regime. First, we discuss the charge state identification and, second, the line identifications. All results are summarized in Tables\,\ref{tab:lines}, \ref{cowan_fit}, and \ref{all_data}.
%
%The simple two-level $^2$D structure of Sn$^{13+}$ in its ground state configuration enables the unambiguous assignment of charge states through the observation of the single bright line connecting these two states. For the Sn$^{12+}$ ion, an energy parameter extrapolation along its isoelectronic sequence \cite{ryabtsev2011resonance} from EUV data is performed, enabling an accurate identification of the observed lines. For the Sn$^{11+}$ and Sn$^{14+}$ ions a combination of predicted transition energies, branching ratios, and Ritz groups results in the identification of most of the observed spectral lines. All results are summarized in Table\,\ref{all_data}.
%
\begin{figure*}
\includegraphics[width=\textwidth]{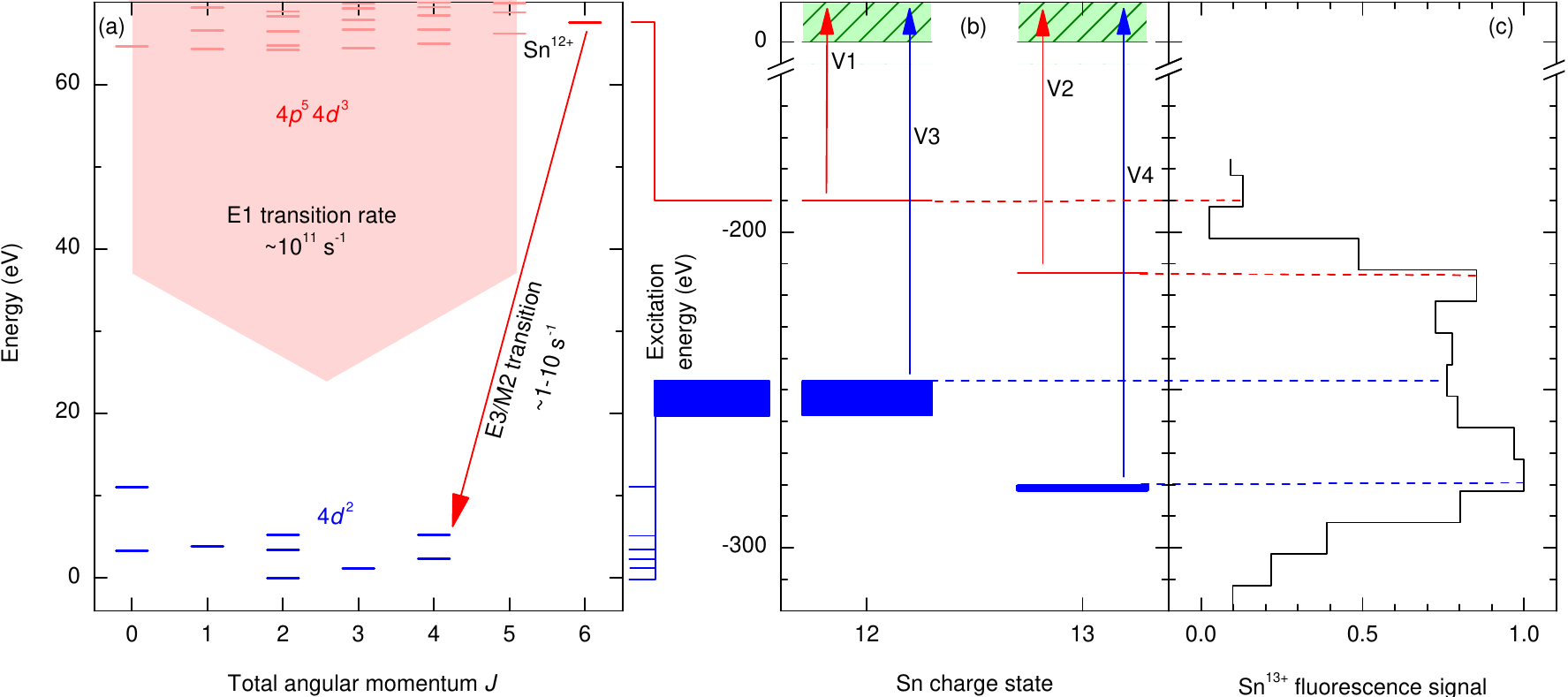}
\caption{Effect of metastable states on the in-EBIT production and fluorescence of Sn$^{13+}$ from the ionization step Sn$^{12+}$ $\rightarrow$ Sn$^{13+}$. (a) Grotrian diagram of Sn$^{12+}$. States in the excited 4$p^5$4$d^{3}$ and 4$p^6$4$d^{1}$4$f^1$ configurations (light red) with total angular momenta $J\leq5$ decay to the ground state configuration 4$p^6$4$d^{2}$ (blue) through fast $E1$ transitions. States with $J\geq6$ are metastable (red) since all possible transitions require a change $\Delta J \geq 2$, and thus accumulate population. (b) The lowest such metastable states for Sn$^{11...14+}$ ions are $\sim$60\,eV closer to the continuum (green) than their ground states (blue). This allows ionization at an energy V1 that is lower than the corresponding ionization potential V3 for the ground state. Between V1 and V3, the value of V2 indicates the ionization threshold for metastable states of the fluorescing Sn$^{13+}$ at which production of Sn$^{14+}$ starts. For the Sn$^{11...14+}$ ions, the difference in (ground state) ionization energies ($\sim$30\,eV) is roughly only half as large as typical metastable-state energies ($\sim$60\,eV). At the threshold V4, Sn$^{13+}$ will be ionized from its ground state, and its population will decrease again. (c) The dependence of the Sn$^{13+}$ fluorescence intensity (black curve) on the electron beam energy qualitatively follows this scenario. The electron beam energy (y-axis in (c)) has been shifted by -20\,eV (compare Fig.\,\ref{fig:optical_zooms}) to account for the electron beam space charge potential.
\label{fig:metastables}}
\end{figure*}%
\begin{figure*}[t]
\includegraphics[width=\textwidth]{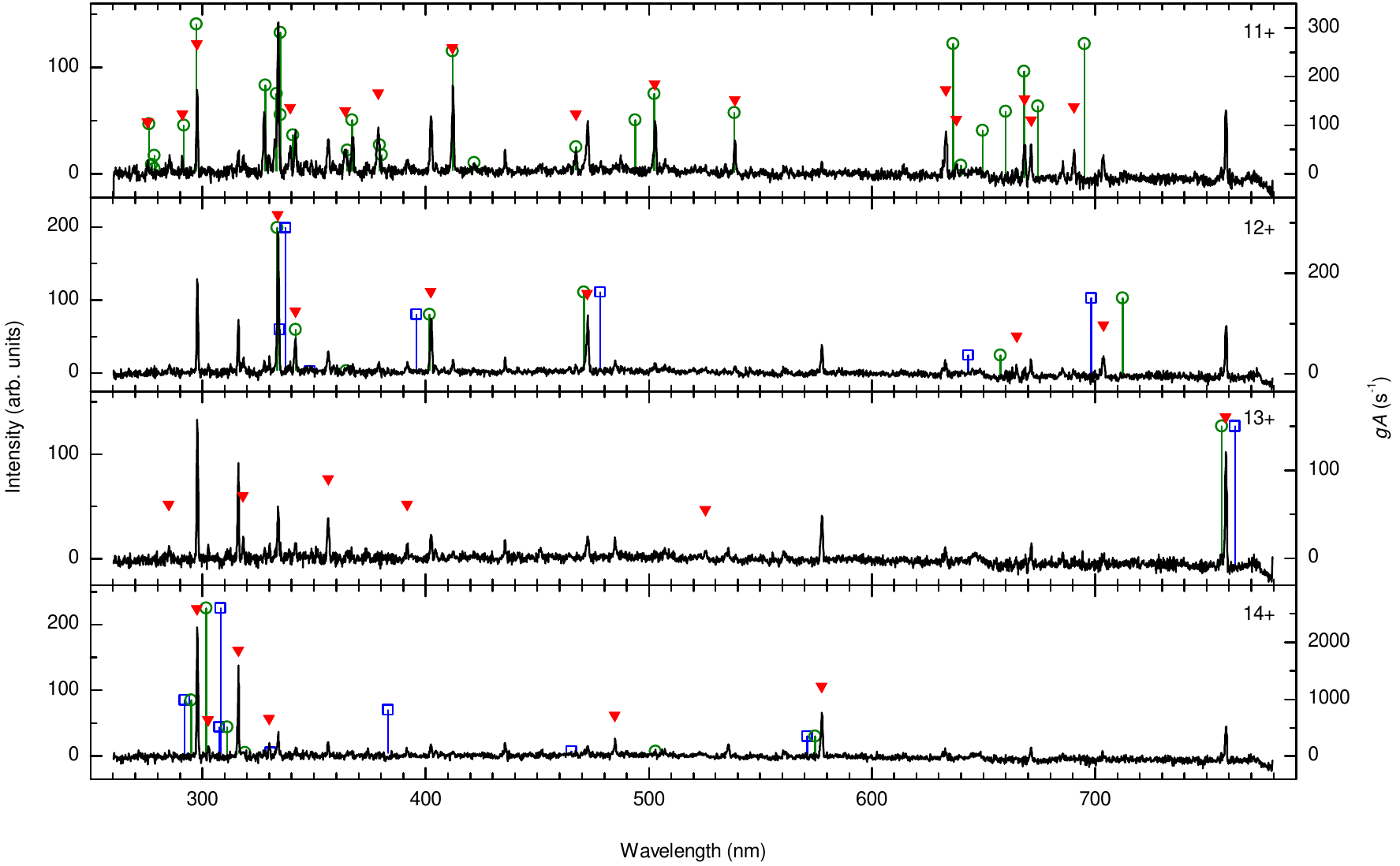}
\caption{Spectra of the charge states Sn$^{11+...14+}$ (orange lines in Fig.\,\ref{fig:Sn_map_charge_states}). Signal intensities here are not corrected for instrument sensitivity. Lines marked with red full triangles belong to the respective charge state. Blue open squares: FSCC predictions. Green circles: semi-empirical Cowan-code calculations (see main text). The weighted transition rates $gA$ are obtained from the Cowan code; they give a measure of the line strength of both theory predictions (right-hand y-axis).
\label{fig:1d_spectra}}
\end{figure*}%
\subsection{Charge state identification}
Scanning the electron beam energy enables the assignment of lines to specific charge states (see Refs.\,\cite{Bekker2015a,lopez2002visible} and references therein) although the absolute determination of a charge number can be challenging (see, e.g., Ref.\,\cite{kobayashi2015extreme}). A typical characteristic of EBIT spectra is that groups of spectral lines that exhibit the same dependence on the acceleration potential can be assigned to the same charge state (see Fig.\,\ref{fig:Sn_map_charge_states}). The appearance and disappearance of sets of lines enables the construction of a ``ladder'' of charge states (see Fig.\,\ref{fig:optical_zooms}). The challenge remaining is to pinpoint a single charge state, with which the others would be easily identified counting up and down. For this purpose, we used the fact that the [Kr]4$d^1$ ground state configuration of Sn$^{13+}$ allows for only a single optical transition originating from the \mbox{4$d$\,\,$^2D_{3/2}$-$^2D_{5/2}$} fine structure splitting (see Fig.\,\ref{fig:sn_level_structure}). This transition is predicted by FSCC to occur near 13\,144\,cm$^{-1}$ (see Table\,\ref{all_data}), in agreement with the 13\,212(25)\,cm$^{-1}$ obtained from the EUV spectra in Refs.\,\cite{churilov2006SnXIII--XV,ryabtsev2008SnXIV}. The brightest line, observed at 758.8(4)\,nm wavelength or 13\,179(8)\,cm$^{-1}$ is an excellent match for it. This unambiguously fixes the identification of Sn$^{13+}$ and with it, that of the other charge states (see Fig.\,\ref{fig:optical_zooms}).

The onsets of fluorescence for all identified charge states occurred at lower acceleration potentials than expected from theory\,\cite{rodrigues2004systematic,NIST_ASD} as is depicted in Fig.\,\ref{fig:optical_zooms}. This, together with the shape of the fluorescence curves, indicates strong contributions from metastable states \cite{borovik2013}. The doubly peaked structure observed for all four charge states hints at metastable states at an excitation energy of $\sim$60\,eV. This excitation energy is similar to that of the low-lying high-$J$ levels of the first excited configuration $4p^5 4d^{m+1}$ that are metastable for decay through electric dipole $E1$ transitions. The observed dependence of the fluorescence of a particular charge state Sn$^{11...14+}$ on the electron beam energy can be qualitatively understood by the sequential opening of four different channels (see Fig.\,\ref{fig:metastables}). Such strong contributions from metastable states are of particular interest for plasma modeling.

A few spectral lines could not be linked to a specific Sn ion. These lines turned out to have only a weak dependence on the electron beam energy. They originate most likely from residual gas or W or Ba ions stemming from the cathode as they remained visible without Sn injection.
%
%A few spectral lines remained visible without Sn injection which could originate from residual gas or W or Ba ions stemming from the cathode. These lines could also be clearly discriminated by their very different, low sensitivity to the EBIT acceleration potential.
%
\subsection{Line identification}
The wavelengths and intensities of the spectral lines for each identified charge state were extracted at the acceleration potential which maximized its fluorescence intensity (see Figs.\,\ref{fig:optical_zooms}, \ref{fig:1d_spectra}). Listed in Table\,\ref{tab:lines} are the centers of the lines, as obtained by fitting Gaussian functions to them. The uncertainty in the determination of the transition wavelengths is dominated by the uncertainty in the spectrometer calibration which is estimated to be $\sim$0.4\,nm. Also listed are the signal intensities given by the area under the fitted Gaussian curves, corrected for the spectral grating efficiency. The accuracy of this procedure is limited, and uncertainties in the determination of the total signal are further introduced by chromatic aberrations of the coupling optics, the finite aperture width, and polarization effects. However, this accuracy should be higher when comparing the relative intensities of transitions with close-lying wavelengths.

Accurate experimentally obtained spectra in the EUV regime are available for the charge states Sn$^{7+}$\,\cite{churilov2006SnVIII}, Sn$^{8+...11+}$\,\cite{churilov2006SnIX--SnXII}, Sn$^{13+}$\,\cite{ryabtsev2008SnXIV}, and Sn$^{12,13,14+}$\,\cite{churilov2006SnXIII--XV}. Transitions in the higher charge states were measured by D'Arcy and coworkers\,\cite{DArcy2009transitions,DArcy2009identification}. We focus our discussion on the charge states Sn$^{11+...14+}$. As pointed out previously \cite{ryabtsev2011resonance}, the identification of weak EUV lines needs to be corrected in previous works. Nevertheless, we will start out by comparing our results to the reference data as is, except in the case of Sn$^{12+}$ for which a new interpretation based on existing EUV spectral data is presented here. Further, a direct comparison is made to the FSCC calculations. In the following the results per charge state are discussed in detail.

\subsubsection{Sn$^{11+}$}
First identifications of transitions within the [Kr]4$d^3$ ground state were obtained from a comparison of the observed energies to the transition energies resulting from the semi-empirical Cowan energy parameters as obtained from Ref.\,\cite{churilov2006SnIX--SnXII}. We associated the predicted transitions of high $gA$ values with the closest-lying, brightest spectral lines (see Fig.\,\ref{fig:1d_spectra}). In this manner, enough levels were identified to enable a fit of the calculations to these levels, improving on the original predictions. This in turn enabled the further identification of observed lines. Iterating the above procedure, all lines attributed to the Sn$^{11+}$  spectrum were identified (see Table\,\ref{tab:lines}). In the final step, the obtained energies of the levels were optimized employing Kramida's code LOPT (for Level Optimization) \cite{kramida2010}. The energy levels thus derived from the experimental wavelengths are collected in Table\,\ref{all_data}. A comparison of the energy parameters is presented in Table\,\ref{cowan_fit}: on the one hand those obtained from \emph{ab initio} HFR calculations, and on the other hand those obtained from fitting the Cowan code to the experimental level values obtained from LOPT. The effective parameters $\beta$ and T1 were fixed at the fitting on the values roughly estimated from the isoelectronic spectrum of Pd$^{7+}$ \cite{ryabtsev2016eighth} and the isonuclear spectrum of Sn$^{7+}$ \cite{Azarov1993}.

Many of the identified levels are connected by Ritz combinations within the experimental uncertainty (red arrows in Fig.\,\ref{fig:sn_level_structure}), enabling the sensitive verification of our $M$1 line identifications. Identifications shown by black arrows in Fig.\,\ref{fig:sn_level_structure} are not supported by Ritz combinations but there is no obvious other choice for their identification. The branching ratios obtained from the signal intensities can be compared to the $gA$ predictions from the Cowan code. Agreement is found within a factor of two except for short wavelength lines at 275.6 and 291.2\,nm, as is to be expected taking into account the experimental uncertainties including those related to the drop of spectrometer efficiency at short wavelengths.

The last column of Table\,\ref{all_data} shows the differences between the level energies obtained in this work and those from the study of the EUV spectrum in Ref.\,\cite{churilov2006SnIX--SnXII}. The magnitude of the differences indicate that the analysis of EUV transitions in Ref.\,\cite{churilov2006SnIX--SnXII} needs to be revised. 

\subsubsection{Sn$^{12+}$} 
We find a very good agreement of our \emph{ab initio} FSCC predictions with the experimental data for the five relatively strong  transitions found in this work. In order to confirm this outcome, we compare them with earlier work. 

All levels of the 4$d^2$ ground configuration of Sn$^{12+}$ were obtained previously in the analysis of EUV lines in discharge spectra \cite{churilov2006SnXIII--XV}. Subsequent studies of the spectra of the Rh$^{7+}$, Pd$^{8+}$, Ag$^{9+}$ and Cd$^{10+}$ ions from the Sn$^{12+}$ isoelectronic sequence and the extrapolation of the isoelectronic regularities to Sn$^{12+}$ showed \cite{ryabtsev2011resonance}, however, that the analysis in Ref.\,\cite{churilov2006SnXIII--XV} of Sn$^{12+}$ should be corrected. 

\begin{figure}[t]
\includegraphics[width=0.48\textwidth]{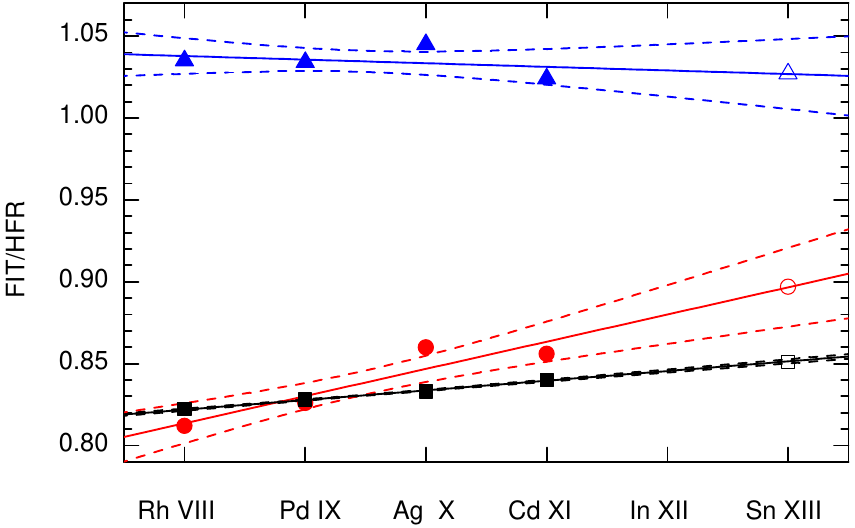}
\caption{Empirical adjustments of scaling factors in the Cowan-code calculations: The ratios (FIT/HFR) for the electrostatic parameters $F^2$ (black squares) and $F^4$ (red circles), and for the spin-orbit parameter $\zeta$ (blue triangles) were obtained by fitting (FIT) to available data \cite{ryabtsev2011resonance} and \emph{ab initio} (HFR) Cowan values of the [Kr]4$d^2$ configuration in Rh$^{7+}$...Cd$^{10+}$. Solid lines represent linear fits and their extrapolation to Sn$^{12+}$ (open symbols). Dashed lines delimit their 1-$\sigma$ confidence bands.
\label{fig:scalingparameters}}
\end{figure}

In Fig.\,\ref{fig:scalingparameters}, the ratios of the energy parameters obtained from fits (FIT) and the \emph{ab initio} Hartree-Fock values (HFR) of the electrostatic and spin-orbit parameters for the [Kr]4$d^2$ configuration in the aforementioned isoelectronic sequence \cite{ryabtsev2011resonance} are shown. 
%These fit values were obtained from the fitting calculations to experimental levels in the framework of the Cowan code with the effective parameter $\alpha$ kept constant at a value of 55 in all spectra. 
The points for Sn$^{12+}$ were obtained by linear extrapolation of the data. The thus obtained scaling factors for the electrostatic parameter $F^4$ and for the spin-orbit parameter, respectively 0.897 and 1.027, are in close agreement with the values 0.900 and 1.030 obtained in Ref.\,\cite{churilov2006SnXIII--XV}. However, the scaling factor for the electrostatic parameter $F^2$, at 0.850, disagrees with the value of 0.829 from Ref.\,\cite{churilov2006SnXIII--XV}. The initial Cowan-code calculations of energy levels, wavelengths, and transition probabilities for $M1$ transitions in this work were performed using the empirical scaling factors obtained from the mentioned extrapolation and taking $\alpha=55$. These prior parameters enabled an accurate prediction of the Sn$^{12+}$ level energies and enabled also the identification of all observed lines for this charge state. This strengthens our assignments of \emph{ab initio} FSCC results to the observed transitions. The branching ratios of the intensity $I$ of the identified transitions $I$(4-6) and $I$(1-6), as well as $I$(1-7) and $I$(2-7) (see Fig.\,\ref{fig:1d_spectra}) can be compared. The experimental ratio $I$(4-6)/$I$(1-6) = 2.6 is in agreement with the number 1.7 from the $gA$ values obtained from Cowan calculations, given the uncertainty in the measurement of the total signal strength. Experimentally, we find $I$(1-7)/$I$(2-7)=1.3, compared to the 1.8 from Cowan theory. One Ritz combination is found, further confirming our identifications of the lines involved. 
%This agreement in the branching ratios is better than that found for the identifications in Sn$^{11+}$, even for the Ritz-confirmed transitions in that charge state. 
%This could in part be due to the much better agreement between theory and experiment regarding the wavelengths of the assigned transitions. 
The energy levels of Sn$^{12+}$ derived from the wavelengths of Table\,\ref{tab:lines} and optimized by LOPT \cite{kramida2010} code are collected in Table\,\ref{all_data}. The magnitude of the differences between the here obtained energy levels and the previously available experimental data \cite{churilov2006SnXIII--XV} indicates that the identification of EUV transitions in that work needs to be revisited.

A comparison of the energy parameters is presented in Table\,\ref{cowan_fit}: those obtained from \emph{ab initio} HFR calculations, and those obtained from the final fitting of the Cowan code to the experimental level values obtained from LOPT. The scaling factors for the Sn$^{12+}$ energy parameters are in agreement with the extrapolated values within the uncertainty of extrapolation. The trends in the change of the scaling factors from Sn$^{12+}$ to Sn$^{11+}$ can be used in extrapolation to the spectrum of Sn$^{10+}$ for a better prediction of its $M$1 transitions; this is part of future work.  

\subsubsection{Sn$^{13+}$}
The \emph{ab initio} FSCC calculations predicting a transition energy of 13\,144\,cm$^{-1}$ for the \mbox{4$d$\,\,$^2D_{3/2}$-$^2D_{5/2}$} fine structure splitting are in good agreement with the line measured at 758.8(4)\,nm wavelength, or 13\,179(8)\,cm$^{-1}$. The data are also in excellent agreement with the result of 13\,212(25)\,cm$^{-1}$ given by \mbox{Refs.\,\cite{ryabtsev2008SnXIV,churilov2006SnXIII--XV}}. This allowed us to use this line for unequivocally identifying the Sn$^{13+}$ charge state. After correcting for the grating efficiency, this line is by far the brightest one observed in our measurements. Features at shorter wavelengths having low intensities stem from transitions between excited, densely packed states within the [Ar]3$d^{10}$4$s^2$4$p^5$4$d^2$ configuration. The identifications of these transitions fall outside the scope of this work.

\subsubsection{Sn$^{14+}$}
The [Kr] ground state configuration 4$p^6$\,\,$^1S_0$ does not exhibit any transitions and all contributions to the optical spectrum have to come from more highly excited configurations like [Ar]3$d^{10}$4$s^2$4$p^5$4$d^1$. All levels of this configuration except the $J=1$ levels are metastable for $E1$ transitions. The metastable states can thus decay and be observed through $M$1 fine structure transitions. 
%Just two $J=1$ energy levels with decaying to the ground level are known from previous studies in the EUV region \cite{sugar1991resonance}. 
Identification of the observed optical lines was performed with aid of the FSCC and Cowan-code calculations. For the latter ones the Slater $F^k$ and $G^k$ parameters were scaled down to 90\% of their \emph{ab initio} values. The spin-orbit parameters for the 4$p$ and 4$d$ electrons were scaled by the factors 1.024 and 1.03, respectively: for the 4$p$ electron we took a value evaluated for the 4$p^5$ configuration in Sn$^{15+}$ \cite{biemont1988}; the value for the 4$d$ electron was extrapolated from the spectra of Sn$^{7+}$ \cite{Azarov1993}, Sn$^{11+}$ (see section above) and Sn$^{12+}$ \cite{churilov2006SnXIII--XV}. Additionally, the average energy of the configuration was adjusted so that the $^3D_1$ value 616\,892\,cm$^{-1}$ obtained experimentally before \cite{sugar1991resonance} was reproduced exactly. The results of the calculations are shown in Table\,\ref{all_data} along with the results of the FSCC calculations. There is good agreement between the wavelengths calculated by these two methods. 
%As FSCC makes no prediction for the transition strengths, 
We used the $gA$ coefficients as obtained from the Cowan code. 

As before, if we associate each predicted transition with the closest-lying, brightest transition, we find that all observed lines can be identified (see Table\,\ref{tab:lines}). Moreover, if we now consider the fact that four of the here identified transitions form a Ritz group (see Fig.\,\ref{fig:sn_level_structure}), we obtain a sensitive tool for checking our tentative assignments. Indeed, the assigned spectral lines form such a group well within the experimental uncertainty. The poor agreement between the experimental and theoretical branching ratio $I$(3-7)/$I$(6-7) could well be related to the poorer agreement between theory and experiment regarding the wavelengths of the assigned transitions. Furthermore, as discussed in the case of Sn$^{11+}$, experimental uncertainties in the determination of the signal intensity increase for short wavelengths like that of the 3-7 transition used in the comparison above.

The identification of the line at 316.3\,nm is achieved considering the branching ratio $I$(3-7)/$I$(4-7)$\approx$2 fits best by assigning this line to the 4-7 transition. It seems that there is no other choice for the weak line at 302.9\,nm but the 8-10 transition even though the $gA$ value for this transition is the highest of all predicted transitions. It should be noted that the other predicted transition with high $gA$, namely 9-11, is absent from the spectrum. Both transitions start from high-lying levels and the apparent low intensities of the corresponding lines could be connected to a lower population of these levels.
The experimental fine structure splittings $E_\textrm{exp}$ of the 4$p^5$4$d^1$ configuration, given in Table\,\ref{all_data}, were obtained inserting the wavelengths of the here identified transitions into LOPT keeping the $^3P_1$ level fixed at the Hartree-Fock value. 
The agreement of \emph{ab initio} FSCC and the semi-empirical Cowan results with experimental data in Table\,\ref{all_data} is excellent. The \emph{ab initio} FSCC method performs at the same level of accuracy as semi-empirical Cowan code that uses extensive input data. Furthermore, comparison of FSCC predictions of the energy of the levels 8 ($^3D_1$ at 617\,515\,cm$^{-1}$) and 12 ($^1P_1$ at 750\,358\,cm$^{-1}$) with previous experimental work in the EUV \cite{sugar1991resonance}, which yielded 616\,892 and 749\,429\,cm$^{-1}$, respectively, is outstanding.\\
%FSCC 750346
%[11] 749429
%
%FSCC 617503
%[11] 616892

Transitions in the EUV stemming from the configurations 4$p^5$4$d^1$-4$p^5$5$p^1$ were observed previously \cite{DArcy2009transitions}. In that work, several peaks in the recorded spectrum were assigned using Cowan code calculations but level energies were not derived because of poor resolution and strong overlap of the various lines. However, there are several cases when two EUV lines starting from the same 4$p^5$5$p^1$ level were measured. Thereby, the separation between the 4$p^5$4$d^1$ levels was found and compared to the levels given in Table\,\ref{all_data}. The splittings $^3P_2$-$^3D_3$ and $^3F_3$-$^1D_2$ are 33\,600(1\,500) and 11\,100(1\,500)\,cm$^{-1}$, respectively. Our values for these intervals, respectively 33\,580(20) and 11\,003(43)\,cm$^{-1}$, are in good agreement. Therefore, the uncertainty of the relative wavelengths in \cite{DArcy2009transitions} might be better than the quoted 0.02\,nm, a finding that also supports our identification of the visible EBIT lines.\\

\section{Conclusions} 
We have re-evaluated the fine structure of Sn$^{11+...14+}$ ions, which are of particular interest for EUV plasma light sources used in next-generation nanolithography. 
%Thereby, we have benchmarked the applicability of \emph{ab initio} Fock space coupled cluster method to such correlated many-electron highly charged ions. 
Experimentally, we combined optical spectroscopy of magnetic dipole $M1$ transitions with charge-state selective ionization in an EBIT. The registered optical spectra were analyzed and line identifications were obtained based on \emph{ab initio} FSCC calculations as well as semi-empirical Cowan code calculations that had adjustable parameters allowing us to fit the observed spectra. Both the FSCC calculations and the semi-empirical Cowan calculations showed a good agreement. The present measurements and identifications provide immediate input for optical plasma-diagnostic tools. Furthermore, our identifications of transitions confirm the very good predictive power of \emph{ab initio} FSCC calculations. Given these encouraging results, it would be particularly advantageous if FSCC could be further developed, as in the current state of the code only atomic systems with a maximum of two open shell electrons/holes can be treated. Comparison of our results with previous work suggests that line identifications based on EUV data need to be revisited. In this type of complex correlated electronic systems, optical spectroscopy delivers data which both complements and challenges studies in the EUV regime and their interpretations.

\begin{table*}[b]
\caption{\label{tab:lines}
Vacuum wavelengths and relative intensities for spectral lines of Sn$^{11+...14+}$ ions measured at the EBIT acceleration potential V$_{\textrm{max}}$ that yielded maximum fluorescence. The experimental wavelength uncertainty of $\sim$0.4\,nm is mainly due to the uncertainty in the calibration. All observed lines could be identified for Sn$^{11,12,14+}$. Our identifications were further confirmed where Ritz combinations could be made (see Fig.\,\ref{fig:sn_level_structure}); contributing lines are indicated by the letters $^{a,b,c,d,e,f}$. Lines near 328\,nm and 368\,nm wavelength, now associated with Sn$^{10+}$, show signs of weak, blended contributions of Sn$^{11+}$, but their unambiguous identification is left for future work. Intensities were derived from the total area of the Gaussian fit, and were corrected for the grating efficiency. Theoretical wavelengths $\lambda_{\mathrm{FSCC}}$  were obtained from FSCC. Wavelengths obtained from semi-empirical Cowan code calculations are marked $\lambda_{\mathrm{Cowan}}$ and the associated weighted transition rates $gA_{ij,\mathrm{Cowan}}$ are given. ``Transition'' gives lower and upper state contributing to transition (from Fig.\,\ref{fig:sn_level_structure}), with their respective configuration and (approximate) term given in the last two columns.}
\begin{tabular}{l c l r r r r c c c }
\hline
\multicolumn{1}{c}{Ion \, \, \, } 					& \multicolumn{1}{c}{V$_{\textrm{max}}$} 	& \multicolumn{1}{c}{$\lambda_{\mathrm{exp}}$}	& \multicolumn{1}{c}{Intensity \, \,}& \multicolumn{1}{c}{$\lambda_{\mathrm{FSCC}}$}& \multicolumn{1}{c}{$\lambda_{\mathrm{Cowan}}$}& \multicolumn{1}{c}{$gA_{ij,\mathrm{Cowan}}$} & Transition & Configuration & Term symbol \\
							&  \multicolumn{1}{c}{(V)}	& \multicolumn{1}{c}{(nm)}											& \multicolumn{1}{c}{(arb. units)}&\multicolumn{1}{c}{(nm)}& \multicolumn{1}{c}{(nm)}& \multicolumn{1}{c}{(s$^{-1}$)} & (see Fig.\,\ref{fig:sn_level_structure})& & \\
\hline	
11+    	& 247   & 275.6$^a$ 		& 8     &      & 276   & 102   & 0-9  &	 [Kr]4$d^3$	& $^4F_{3/2}$-$^2D_{3/2}$	\\
				&				& 291.2$^{b,c}$	& 14    &      & 292   & 100   & 2-13 &							& $^4F_{7/2}$-$^2G_{9/2}$	\\
				&       & 297.8  		& 23    &      & 297   & 308   & 2-12 &	 						& $^4F_{7/2}$-$^2D_{5/2}$	\\
%				&       & 328?  		&     &        &       &       &      &	 						& $^4F_{7/2}$-$^2D_{5/2}$	\\
				&       & 339.5$^d$			& 34    &      & 341   & 79    & 3-13 &	 						&	$^4F_{9/2}$-$^2G_{9/2}$	\\
				&       & 364.2 				& 44    &      & 365   & 48    & 6-16 &	 						&	$^2G_{7/2}$-$^2F_{5/2}$	\\
				&       & 378.9  		& 81    &      & 379   & 59    & 3-11 &	 						& $^4F_{9/2}$-$^2H_{11/2}$ \\
				&       & 412.4  		& 136   &      & 412   & 253   & 1-6  &	 						&	$^4F_{5/2}$-$^2G_{7/2}$ \\
				&       & 467.5$^a$ 		& 39    &      & 468   & 55    & 0-4  &	 						& $^4F_{3/2}$-$^4P_{3/2}$ \\
				&       & 503.0$^c$ 		& 129   &      & 503   & 165   & 2-8	 &	 					& $^4F_{7/2}$-$^2H_{9/2}$ \\
				&       & 538.7$^b$ 		& 75    &      & 539   & 125   & 2-6	 &	 					& $^4F_{7/2}$-$^2G_{7/2}$ \\
				&       & 633.3$^b$ 		& 178   &      & 637   & 268   & 6-13 &	 						& $^2G_{7/2}$-$^2G_{9/2}$ \\
				&       & 638.1  		& 53    &      & 640   & 17    & 1-4  &	 						& $^4F_{5/2}$-$^4P_{3/2}$ \\
				&       & 668.5$^d$ 		& 143   &      & 668   & 211   & 3-8  &	 						& $^4F_{9/2}$-$^2H_{9/2}$ \\
				&       & 671.5$^a$	 		&	 51   &      & 674   & 139   & 4-9  &	 						&	$^4P_{3/2}$-$^2D_{3/2}$	\\\vspace{2mm}
				&       & 690.7$^{c,d}$ &   92  &      & 695   & 267   & 8-13 &	 						& $^2H_{9/2}$-$^2G_{9/2}$ \\
12+    	& 267   & 334.1 				& 247   & 337  & 334   & 290   & 1-7  &[Kr]4$d^2$		& $^3F_{3}$-$^1G_{4}$		 \\
				&				& 341.8$^e$ 				& 65    & 335  & 342   & 87    & 1-6  &							& $^3F_{3}$-$^3P_{2}$		 \\
				&       & 402.6 				& 122   & 396  & 402   & 117   & 0-4  &	 						& $^3F_{2}$-$^1D_{2}$	\\
				&       & 472.7 				& 185   & 478  & 471   & 162   & 2-7  &	 
										& $^3F_{4}$-$^1G_{4}$	\\
&       & 665.0$^e$                                                            & 63    & 643  & 658          &            36   & 1-4  &                        & $^3F_{3}$-$^1D_{2}$                \\\vspace{2mm}														&       & 703.9$^e$ 				& 167   & 698  & 712   & 151   & 4-6  &	 						& $^1D_{2}$-$^3P_{2}$	\\
13+    	& 297   & 		285.2 		& 11    &      &       &  			&			&								&					\\
				&				& 318.5 				& 19    &      &       &  			&			&								&					\\
				&       & 356.5 				& 62    &      &       &  			&			&	 							&					\\
				&       & 391.9 				& 17    &      &       &  			&			&	 							&					\\
				&       & 525.6 				& 24    &      &       &  			&			&	 							&					\\\vspace{2mm}
				&       & 758.8 				& 518   & 763  & 757   &  	 150		& 0-1	&[Kr]4$d^1$			&	$^2D_{3/2}$-$^2D_{5/2}$	\\
14+				& 327   		& 297.9$^f$		& 211 		& 292 &	 295			&		984		& 3-7  &4$p^5$ 4$d^1$				&	$^3P_2$-$^3D_3$	\\
				&				& 302.9 		  & 15  	& 308 &	 302			&		2597		& 8-10 &							& $^3D_1$-$^3D_2$		\\
				&				& 316.3 		  & 109 	& 308 &	 311			&		516		& 4-7  &	 						&	$^3F_3$-$^3D_3$	\\
				&				& 330.2$^f$		  & 20  	& 331 &	 319			&		62		& 2-6  &	 						&	$^3P_1$-$^1D_2$	\\
				&				& 485.1$^f$		  & 51  	& 465 &	 503			&		80		& 6-7  &	 						&	$^1D_2$-$^3D_3$	\\
				&				& 577.7$^f$		  & 207 	& 571 &	 575			&		349		& 2-3  &	 						&	$^3P_1$-$^3P_2$	\\
\hline
\end{tabular}
\end{table*}

\begin{table*}[]
\centering
\caption{Cowan code Hartree-Fock (HFR) and fitted (FIT) parameters (with uncertainties $\delta_{\mathrm{fit}}$), their ratios (scaling factors), and root mean square deviations $\sigma$ of the fits as calculated for the $4d^k + 4d^{k-1} 5s + 4d^{k-2} 5s^2$ interacting configurations ($k=3, 2$ respectively in the Sn$^{11+}$ and Sn$^{12+}$ spectra). The electrostatic parameters were scaled by a factor 0.85 whereas average energies and spin-orbit parameters were not scaled in the unknown $4d^{k-1} 5s$ and $4d^{k-2} 5s^2$ configurations. All parameters are given in units of cm$^{-1}$.}
\label{cowan_fit}
\begin{tabular}{l r r r r r r r r}
\hline
\multicolumn{1}{c}{Parameter} & \multicolumn{4}{c}{Sn$^{11+}$} & \multicolumn{4}{c}{Sn$^{12+}$}  \\
     & \multicolumn{1}{c}{HFR} & \multicolumn{1}{c}{FIT} & \multicolumn{1}{c}{$\delta_{\mathrm{fit}}$}  & \multicolumn{1}{c}{FIT/HFR} & \multicolumn{1}{c}{HFR} & \multicolumn{1}{c}{FIT} & \multicolumn{1}{c}{$\delta_{\mathrm{fit}}$}  & \multicolumn{1}{c}{FIT/HFR} \\ \hline
$E_{av}$			 & 41\,025 	& 37\,533   &     17	&       &  26\,035	& 24\,435 &     2 & 		  \\
$F^2(4d,4d)$   & 114\,058 & 97\,715   &    121	& 0.857 & 116\,377	& 99\,790 &    21 & 0.857 \\
$F^4(4d,4d)$   & 76\,975 	& 68\,047   &    323	& 0.884 & 78\,692	& 69\,026 &   270 & 0.877 \\
$\alpha$   		 &     - 	&    64   &      3	&       &       -	&    68 &     1 & 			\\
$\beta$        &     - 	&  -600   & fixed		&       &       -	&  -600 & fixed & 			\\
$T1$  			   &     - 	&    -4.4 & fixed 	&       &       -	&       &       & 			\\
$\zeta(4d)$    & 4\,638 	&  4\,774   &     16	& 1.029 &   4\,868	&  5\,028 &     2 & 1.033 \\
$\sigma$   		 &        &     57  & 			  & 		  &        	&     4 &       &  			\\\hline
\end{tabular}                                              
\end{table*}

\begin{table*}
\centering
\caption{Energy levels $E_\mathrm{exp}$ (all energies in cm$^{-1}$) derived from the experimental data using Kramida's LOPT algorithm \cite{kramida2010}, \emph{ab initio} FSCC calculations $E_{\mathrm{FSCC}}$ (including individual contributions from the Breit interaction $\Delta E_\mathrm{Breit}$ and lamb-shift $\Delta E_\mathrm{LS}$) as well as semi-empirical Cowan code calculations $E_\mathrm{Cowan}$ of the investigated fine-structure configurations in Sn$^{11+...14+}$. Levels (from Fig.\,\ref{fig:sn_level_structure}) are ordered by their energies; we use \emph{LS}-term notations as approximated from the Cowan code. The dispersive energy uncertainty $D_1$ is the uncertainty relative to any other term, and the energy uncertainty $D_2$ is that relative to ground level (to the $^3P_1$ level in Sn$^{14+}$ which is offset to zero $E_{\mathrm{FSCC}}$ and $E_\mathrm{Cowan}$ in this Table; $\Delta E_\mathrm{Breit}$ and $\Delta E_\mathrm{LS}$ are given with respect to the ground level $^1S_{0}$), following the definition in \cite{kramida2010}. The number of spectral lines used for the determination of each level energy is given by $N$. Semi-empirical Cowan code calculations are given in column $E_\mathrm{Cowan}$. The identification of spectral lines in Sn$^{11+}$ and Sn$^{12+}$ used an iterative fit procedure based on the measured spectra (see Section\,\ref{Cowan}, Table\,\ref{cowan_fit}); for Sn$^{13+}$ the value for $E_\mathrm{Cowan}$ was obtained from single configuration HFR; and for Sn$^{14+}$ the ratio FIT/HFR of Cowan parameters was determined by isoelectronic extrapolations and data from \cite{sugar1991resonance} (see main text). Energies deduced from vacuum-spark EUV spectra $E_{\mathrm{vs}}$ have uncertainties of $\sim$40\,cm$^{-1}$ \cite{churilov2006SnIX--SnXII,churilov2006SnXIII--XV}. The difference $\Delta E_{\mathrm{vs}}=E_\mathrm{exp}-E_{\mathrm{vs}}$ of that interpretation with our own identifications suggests the need for revision of those earlier identifications.}
\label{all_data}
\begin{tabular}{l c c r r r r r r r r r r}
\hline
\multicolumn{1}{c}{Ion} & \multicolumn{1}{c}{Level} & \multicolumn{1}{c}{Term } & \multicolumn{4}{c}{Experiment}  & \multicolumn{3}{c}{FSCC} &  \multicolumn{1}{c}{$E_\mathrm{Cowan}$}& \multicolumn{1}{c}{\, \, \,$E_{\mathrm{vs}}$ \,} & \multicolumn{1}{c}{$\Delta E_{\mathrm{vs}}$} \\
 &      &           &  \multicolumn{1}{c}{$E_\mathrm{exp}$} & \multicolumn{1}{c}{$D_1$}     & \multicolumn{1}{c}{$D_2$}    & \multicolumn{1}{c}{$N$} &\multicolumn{1}{c}{$\Delta E_\mathrm{Breit}$} & \multicolumn{1}{c}{$\Delta E_\mathrm{LS}$}&\multicolumn{1}{c}{$E_{\mathrm{FSCC}}$}    &&&\\ \hline
11+ 4$d^3$		& 0  & $^4F_{3/2}$  & 0 			 	& 17 & 0  & 2 &       &    &         &	    0		  & 0 		& 0			\\
							& 1  & $^4F_{5/2}$  & 5\,719 		 	& 9 & 20 & 2 &       &    &         &	 5\,758		  & 5\,760  & -36 	\\
							& 2  & $^4F_{7/2}$  & 11\,403 	 	& 10 & 30 & 3 &       &    &         &	11\,445		  & 11\,465 & -57 	\\
							& 3  & $^4F_{9/2}$  & 16\,321 	 	& 9  & 30 & 2 &       &    &         &	16\,375		  & 16\,390 & -64 	\\
							& 4  & $^4P_{3/2}$  & 21\,391 	 	& 6  & 17 & 3 &       &    &         &	21\,390		  & 20\,486 & 905 	\\
							& 5  & $^4P_{1/2}$  &   			 	& 	 &    &   &       &    &         &	23\,341		  &  		 	&     	\\
							& 6  & $^2G_{7/2}$  & 29\,967 	 	& 8  & 30 & 3 &       &    &         &	30\,011		  & 30\,057 & -84 	\\
							& 7  & $^4P_{5/2}$  &   	 		 	& 	 &  	&   &       &    &         &	31\,130		  & 			&  	 		\\
							& 8  & $^2H_{9/2}$  & 31\,280 	 	& 6  & 30 & 3 &       &    &         &	31\,337		  & 31\,110 & 175 	\\
							& 9  & $^2D_{3/2}$  & 36\,283 	 	& 9 & 19 & 2 &       &    &         &	36\,216		  & 35\,810 & 471 	\\
							& 10 & $^4P_{1/2}$  &   	 		 	&    &    & 	&       &    &         &	38\,736		  & 	 		&     	\\
							& 11 & $^2H_{11/2}$ & 42\,713 	 	& 28 & 44 & 1 &       &    &         &	42\,726		  & 42\,230 & 488 	\\
							& 12 & $^2D_{5/2}$  & 44\,983 	 	& 45 & 56 & 1 &       &    &         &	45\,081		  & 44\,990 & -2  	\\
							& 13 & $^2G_{9/2}$  & 45\,759 	 	& 6  & 30 & 4 &       &    &         &	45\,720		  & 45\,705 & 59  	\\
							& 14 & $^2P_{3/2}$	&   	 		 	&    &  	&   &       &    &         &	51\,365		  &   		&     	\\
							& 15 & $^2F_{7/2}$  & 	 			& 	 & 	 & 	 &       &    &          & 	57\,261		  & 55\,460 & 	\\\vspace{2mm}
%							& 15 & $^2F_{7/2}$  & 57\,170?	 	& 30 & 43 & 1 &       &    &         &	57\,261		  & 55\,460 & 1\,710	\\\vspace{2mm}
							& 16 & $^2F_{5/2}$  & 57\,425 	 	& 30 & 43 & 1 &       &    &         &	57\,414		  & 55\,660 & 1\,770	\\
12+ 4$d^2$ 		& 0  & $^3F_{2  }$  & 0 			 	& 25 & 0  & 1 &   0    & 0  & 0       &	    0		  & 0 		& 	 		\\
							& 1  & $^3F_{3  }$  & 9\,786 		 	& 9  & 30 & 2 &  -374     & 29 & 9\,738    &	 9\,780		  & 9\,745  & 41  	\\
							& 2  & $^3F_{4  }$  & 18\,564 	 	& 18 & 50 & 1 &   -655    & 57 & 18\,507   &	18\,563		  & 18\,480 & 84  	\\
							& 3  & $^3P_{0  }$	&   	 		 	& 	 &  	&  	&  -87     & 0  & 23\,642   &	22\,649		  & 			&     	\\
							& 4  & $^1D_{2  }$  & 24\,838 	 	& 6  & 25 & 2 &   -355    & 29 & 25\,285   &	24\,835		  & 24\,320 & 518 	\\
							& 5  & $^3P_{1  }$	&   	 		 	& 	 &  	&   &  -238     & 29 & 28\,750   &	27\,905		  & 			&     	\\
							& 6  & $^3P_{2  }$  & 39\,044 	 	& 8  & 30 & 2 &   -425    & 57 & 39\,636   &	39\,042		  & 38\,370 & 674 	\\
							& 7  & $^1G_{4  }$  & 39\,718 	 	& 36 & 44 & 1 &   -983    & 29 & 39\,381   &	39\,715		  & 38\,830 & 888 	\\\vspace{2mm}
							& 8  & $^1S_{0  }$  &   	 		 	& 	 &  	& 	& -381      & 57 & 83\,202   &	80\,700		  & 			&     	\\
13+ 4$d^1$		& 0  &  $^2D_{3/2}$ & 0    		  &    &    &   & 0     & 0  & 0       & 0					  &	0		 	& 0	 		\\\vspace{2mm}
							& 1  &  $^2D_{5/2}$ & 13\,179  			&		 &    &	  & -439  & 30 & 13\,144   &	12\,740			& 	13\,212		&   -33  	\\ 
14+ $4p^6$					& 0  &  $^1S_{0}$	&   	 		 	& 	 &    	& 	&    0    &  0 & -540\,785 &	-539\,447	  &			  &     	\\
4$p^5$4$d^1$  				& 1  & $^3P_0$      &   	 		 	& 	 &    	& 	& -870    & -128  & -8\,962   &	-8\,970		  &			  &     	\\
							& 2  & $^3P_1$      & 0 				& 11 & 0  	& 2 & -1\,054 & -128  & 0 		   &	0				  &			  & 			\\
							& 3  & $^3P_2$      & 17\,311 			& 12 & 12	& 2 & -1\,262 &  -97  & 17\,544   &	17\,392		  &       &     	\\
							& 4  & $^3F_3$      & 19\,275 			& 40 & 50	& 1 & -1\,048 & -128  & 19\,247   &	19\,115		  &       &     	\\
							& 5  & $^3F_4$     	&  					& 	 &  	& 	& -1\,464 &  -97  & 21\,027   &	20\,991		  & 			&     	\\
							& 6  & $^1D_2$      & 30\,278 			& 16 & 30	& 2 & -1\,197 & -128  & 30\,252   &	31\,345		  &       &     	\\
							& 7  & $^3D_3$      & 50\,891 			& 16 & 30	& 2 & -1\,410 &  -97   & 51\,770   &	51\,208		  &       &     	\\
							& 8  & $^3D_1$     	&   	 		 	& 	 &  	& 	& -1\,535 &   12 & 76\,730   &	77\,445		  &			 	&     	\\
							& 9  & $^3F_2$     	&   	 		 	& 	 &  	& 	& -1\,996 &   12 & 91\,543   &	93\,484		  &			 	&     	\\		
							& 10 & $^3D_2$     	&   	 		 	& 	 &  	& 	& -2\,348 &   46 & 109\,202  &	110\,507	  &			 	&     	\\
							& 11 & $^1F_3$     	&   	 		 	& 	 &  	& 	& -2\,445 &   46 & 117\,668  &	118\,362	  &			 	&     	\\
							& 12 & $^1P_1$     	&   	 		 	& 	 &  	& 	& -1\,853 &   12 & 209\,573  &	222\,563	  &			 	&     	\\\hline
\end{tabular}
\end{table*}

\bibliography{library_vclean_v13}

\begin{thebibliography}{42}%
\makeatletter
\providecommand \@ifxundefined [1]{%
 \@ifx{#1\undefined}
}%
\providecommand \@ifnum [1]{%
 \ifnum #1\expandafter \@firstoftwo
 \else \expandafter \@secondoftwo
 \fi
}%
\providecommand \@ifx [1]{%
 \ifx #1\expandafter \@firstoftwo
 \else \expandafter \@secondoftwo
 \fi
}%
\providecommand \natexlab [1]{#1}%
\providecommand \enquote  [1]{``#1''}%
\providecommand \bibnamefont  [1]{#1}%
\providecommand \bibfnamefont [1]{#1}%
\providecommand \citenamefont [1]{#1}%
\providecommand \href@noop [0]{\@secondoftwo}%
\providecommand \href [0]{\begingroup \@sanitize@url \@href}%
\providecommand \@href[1]{\@@startlink{#1}\@@href}%
\providecommand \@@href[1]{\endgroup#1\@@endlink}%
\providecommand \@sanitize@url [0]{\catcode `\\12\catcode `\$12\catcode
  `\&12\catcode `\#12\catcode `\^12\catcode `\_12\catcode `\%12\relax}%
\providecommand \@@startlink[1]{}%
\providecommand \@@endlink[0]{}%
\providecommand \url  [0]{\begingroup\@sanitize@url \@url }%
\providecommand \@url [1]{\endgroup\@href {#1}{\urlprefix }}%
\providecommand \urlprefix  [0]{URL }%
\providecommand \Eprint [0]{\href }%
\providecommand \doibase [0]{http://dx.doi.org/}%
\providecommand \selectlanguage [0]{\@gobble}%
\providecommand \bibinfo  [0]{\@secondoftwo}%
\providecommand \bibfield  [0]{\@secondoftwo}%
\providecommand \translation [1]{[#1]}%
\providecommand \BibitemOpen [0]{}%
\providecommand \bibitemStop [0]{}%
\providecommand \bibitemNoStop [0]{.\EOS\space}%
\providecommand \EOS [0]{\spacefactor3000\relax}%
\providecommand \BibitemShut  [1]{\csname bibitem#1\endcsname}%
\let\auto@bib@innerbib\@empty
%</preamble>
\bibitem [{\citenamefont {Benschop}\ \emph {et~al.}(2008)\citenamefont
  {Benschop}, \citenamefont {Banine}, \citenamefont {Lok},\ and\ \citenamefont
  {Loopstra}}]{Benschop2008}%
  \BibitemOpen
  \bibfield  {author} {\bibinfo {author} {\bibfnamefont {J.}~\bibnamefont
  {Benschop}}, \bibinfo {author} {\bibfnamefont {V.}~\bibnamefont {Banine}},
  \bibinfo {author} {\bibfnamefont {S.}~\bibnamefont {Lok}}, \ and\ \bibinfo
  {author} {\bibfnamefont {E.}~\bibnamefont {Loopstra}},\ }\href@noop {}
  {\bibfield  {journal} {\bibinfo  {journal} {J. Vac. Sci. Technol. B}\
  }\textbf {\bibinfo {volume} {26}},\ \bibinfo {pages} {2204} (\bibinfo {year}
  {2008})}\BibitemShut {NoStop}%
\bibitem [{\citenamefont {Banine}\ \emph {et~al.}(2011)\citenamefont {Banine},
  \citenamefont {Koshelev},\ and\ \citenamefont {Swinkels}}]{Banine2011}%
  \BibitemOpen
  \bibfield  {author} {\bibinfo {author} {\bibfnamefont {V.~Y.}\ \bibnamefont
  {Banine}}, \bibinfo {author} {\bibfnamefont {K.~N.}\ \bibnamefont
  {Koshelev}}, \ and\ \bibinfo {author} {\bibfnamefont {G.~H. P.~M.}\
  \bibnamefont {Swinkels}},\ }\href@noop {} {\bibfield  {journal} {\bibinfo
  {journal} {J. Phys. D: Appl. Phys.}\ }\textbf {\bibinfo {volume} {44}},\
  \bibinfo {pages} {253001} (\bibinfo {year} {2011})}\BibitemShut {NoStop}%
\bibitem [{\citenamefont {O'Sullivan}\ \emph {et~al.}(2015)\citenamefont
  {O'Sullivan}, \citenamefont {Li}, \citenamefont {D'Arcy}, \citenamefont
  {Dunne}, \citenamefont {Hayden}, \citenamefont {Kilbane}, \citenamefont
  {McCormack}, \citenamefont {Ohashi}, \citenamefont {O'Reilly}, \citenamefont
  {Sheridan}, \citenamefont {Sokell}, \citenamefont {Suzuki},\ and\
  \citenamefont {Higashiguchi}}]{OSullivan2015}%
  \BibitemOpen
  \bibfield  {author} {\bibinfo {author} {\bibfnamefont {G.}~\bibnamefont
  {O'Sullivan}}, \bibinfo {author} {\bibfnamefont {B.}~\bibnamefont {Li}},
  \bibinfo {author} {\bibfnamefont {R.}~\bibnamefont {D'Arcy}}, \bibinfo
  {author} {\bibfnamefont {P.}~\bibnamefont {Dunne}}, \bibinfo {author}
  {\bibfnamefont {P.}~\bibnamefont {Hayden}}, \bibinfo {author} {\bibfnamefont
  {D.}~\bibnamefont {Kilbane}}, \bibinfo {author} {\bibfnamefont
  {T.}~\bibnamefont {McCormack}}, \bibinfo {author} {\bibfnamefont
  {H.}~\bibnamefont {Ohashi}}, \bibinfo {author} {\bibfnamefont
  {F.}~\bibnamefont {O'Reilly}}, \bibinfo {author} {\bibfnamefont
  {P.}~\bibnamefont {Sheridan}}, \bibinfo {author} {\bibfnamefont
  {E.}~\bibnamefont {Sokell}}, \bibinfo {author} {\bibfnamefont
  {C.}~\bibnamefont {Suzuki}}, \ and\ \bibinfo {author} {\bibfnamefont
  {T.}~\bibnamefont {Higashiguchi}},\ }\href@noop {} {\bibfield  {journal}
  {\bibinfo  {journal} {J. Phys. B}\ }\textbf {\bibinfo {volume} {48}},\
  \bibinfo {pages} {144025} (\bibinfo {year} {2015})}\BibitemShut {NoStop}%
\bibitem [{\citenamefont {Azarov}\ and\ \citenamefont
  {Joshi}(1993)}]{Azarov1993}%
  \BibitemOpen
  \bibfield  {author} {\bibinfo {author} {\bibfnamefont {V.~I.}\ \bibnamefont
  {Azarov}}\ and\ \bibinfo {author} {\bibfnamefont {Y.~N.}\ \bibnamefont
  {Joshi}},\ }\href
  {http://iopscience.iop.org/article/10.1088/0953-4075/26/20/011} {\bibfield
  {journal} {\bibinfo  {journal} {J. Phys. B}\ }\textbf {\bibinfo {volume}
  {26}},\ \bibinfo {pages} {3495} (\bibinfo {year} {1993})}\BibitemShut
  {NoStop}%
\bibitem [{\citenamefont {Tolstikhina}\ \emph {et~al.}(2006)\citenamefont
  {Tolstikhina}, \citenamefont {Churilov}, \citenamefont {Ryabtsev},\ and\
  \citenamefont {Koshelev}}]{tolstikhina2006ATOMICDATA}%
  \BibitemOpen
  \bibfield  {author} {\bibinfo {author} {\bibfnamefont {I.~Y.}\ \bibnamefont
  {Tolstikhina}}, \bibinfo {author} {\bibfnamefont {S.~S.}\ \bibnamefont
  {Churilov}}, \bibinfo {author} {\bibfnamefont {A.~N.}\ \bibnamefont
  {Ryabtsev}}, \ and\ \bibinfo {author} {\bibfnamefont {K.~N.}\ \bibnamefont
  {Koshelev}},\ }in\ \href@noop {} {\emph {\bibinfo {booktitle} {EUV sources
  for lithography}}},\ Vol.\ \bibinfo {volume} {149},\ \bibinfo {editor}
  {edited by\ \bibinfo {editor} {\bibfnamefont {V.}~\bibnamefont {Bakshi}}}\
  (\bibinfo  {publisher} {SPIE Press},\ \bibinfo {year} {2006})\ p.\ \bibinfo
  {pages} {113}\BibitemShut {NoStop}%
\bibitem [{\citenamefont {Churilov}\ and\ \citenamefont
  {Ryabtsev}(2006{\natexlab{a}})}]{churilov2006SnVIII}%
  \BibitemOpen
  \bibfield  {author} {\bibinfo {author} {\bibfnamefont {S.~S.}\ \bibnamefont
  {Churilov}}\ and\ \bibinfo {author} {\bibfnamefont {A.~N.}\ \bibnamefont
  {Ryabtsev}},\ }\href@noop {} {\bibfield  {journal} {\bibinfo  {journal} {Opt.
  Spectrosc.}\ }\textbf {\bibinfo {volume} {100}},\ \bibinfo {pages} {660}
  (\bibinfo {year} {2006}{\natexlab{a}})}\BibitemShut {NoStop}%
\bibitem [{\citenamefont {Churilov}\ and\ \citenamefont
  {Ryabtsev}(2006{\natexlab{b}})}]{churilov2006SnXIII--XV}%
  \BibitemOpen
  \bibfield  {author} {\bibinfo {author} {\bibfnamefont {S.~S.}\ \bibnamefont
  {Churilov}}\ and\ \bibinfo {author} {\bibfnamefont {A.~N.}\ \bibnamefont
  {Ryabtsev}},\ }\href@noop {} {\bibfield  {journal} {\bibinfo  {journal} {Opt.
  Spectrosc.}\ }\textbf {\bibinfo {volume} {101}},\ \bibinfo {pages} {169}
  (\bibinfo {year} {2006}{\natexlab{b}})}\BibitemShut {NoStop}%
\bibitem [{\citenamefont {Ryabtsev}\ \emph {et~al.}(2008)\citenamefont
  {Ryabtsev}, \citenamefont {Kononov},\ and\ \citenamefont
  {Churilov}}]{ryabtsev2008SnXIV}%
  \BibitemOpen
  \bibfield  {author} {\bibinfo {author} {\bibfnamefont {A.~N.}\ \bibnamefont
  {Ryabtsev}}, \bibinfo {author} {\bibfnamefont {{\'E}.~Y.}\ \bibnamefont
  {Kononov}}, \ and\ \bibinfo {author} {\bibfnamefont {S.~S.}\ \bibnamefont
  {Churilov}},\ }\href@noop {} {\bibfield  {journal} {\bibinfo  {journal} {Opt.
  Spectrosc.}\ }\textbf {\bibinfo {volume} {105}},\ \bibinfo {pages} {844}
  (\bibinfo {year} {2008})}\BibitemShut {NoStop}%
\bibitem [{\citenamefont {Churilov}\ and\ \citenamefont
  {Ryabtsev}(2006{\natexlab{c}})}]{churilov2006SnIX--SnXII}%
  \BibitemOpen
  \bibfield  {author} {\bibinfo {author} {\bibfnamefont {S.~S.}\ \bibnamefont
  {Churilov}}\ and\ \bibinfo {author} {\bibfnamefont {A.~N.}\ \bibnamefont
  {Ryabtsev}},\ }\href@noop {} {\bibfield  {journal} {\bibinfo  {journal}
  {Phys. Scr.}\ }\textbf {\bibinfo {volume} {73}},\ \bibinfo {pages} {614}
  (\bibinfo {year} {2006}{\natexlab{c}})}\BibitemShut {NoStop}%
\bibitem [{\citenamefont {Svendsen}\ and\ \citenamefont
  {O'Sullivan}(1994)}]{Svendsen1994}%
  \BibitemOpen
  \bibfield  {author} {\bibinfo {author} {\bibfnamefont {W.}~\bibnamefont
  {Svendsen}}\ and\ \bibinfo {author} {\bibfnamefont {G.}~\bibnamefont
  {O'Sullivan}},\ }\href {\doibase 10.1103/PhysRevA.50.3710} {\bibfield
  {journal} {\bibinfo  {journal} {Phys. Rev. A}\ }\textbf {\bibinfo {volume}
  {50}},\ \bibinfo {pages} {3710} (\bibinfo {year} {1994})}\BibitemShut
  {NoStop}%
\bibitem [{\citenamefont {Sugar}\ \emph {et~al.}(1991)\citenamefont {Sugar},
  \citenamefont {Rowan},\ and\ \citenamefont {Kaufman}}]{sugar1991resonance}%
  \BibitemOpen
  \bibfield  {author} {\bibinfo {author} {\bibfnamefont {J.}~\bibnamefont
  {Sugar}}, \bibinfo {author} {\bibfnamefont {W.~L.}\ \bibnamefont {Rowan}}, \
  and\ \bibinfo {author} {\bibfnamefont {V.}~\bibnamefont {Kaufman}},\
  }\href@noop {} {\bibfield  {journal} {\bibinfo  {journal} {JOSA B}\ }\textbf
  {\bibinfo {volume} {8}},\ \bibinfo {pages} {2026} (\bibinfo {year}
  {1991})}\BibitemShut {NoStop}%
\bibitem [{\citenamefont {Sugar}\ \emph {et~al.}(1992)\citenamefont {Sugar},
  \citenamefont {Rowan},\ and\ \citenamefont {Kaufman}}]{sugar1992rb}%
  \BibitemOpen
  \bibfield  {author} {\bibinfo {author} {\bibfnamefont {J.}~\bibnamefont
  {Sugar}}, \bibinfo {author} {\bibfnamefont {W.~L.}\ \bibnamefont {Rowan}}, \
  and\ \bibinfo {author} {\bibfnamefont {V.}~\bibnamefont {Kaufman}},\
  }\href@noop {} {\bibfield  {journal} {\bibinfo  {journal} {JOSA B}\ }\textbf
  {\bibinfo {volume} {9}},\ \bibinfo {pages} {1959} (\bibinfo {year}
  {1992})}\BibitemShut {NoStop}%
\bibitem [{\citenamefont {Ohashi}\ \emph {et~al.}(2009)\citenamefont {Ohashi},
  \citenamefont {Suda}, \citenamefont {Tanuma}, \citenamefont {Fujioka},
  \citenamefont {Nishimura}, \citenamefont {Nishihara}, \citenamefont {Kai},
  \citenamefont {Sasaki}, \citenamefont {Sakaue}, \citenamefont {Nakamura}
  \emph {et~al.}}]{ohashi2009complementary}%
  \BibitemOpen
  \bibfield  {author} {\bibinfo {author} {\bibfnamefont {H.}~\bibnamefont
  {Ohashi}}, \bibinfo {author} {\bibfnamefont {S.}~\bibnamefont {Suda}},
  \bibinfo {author} {\bibfnamefont {H.}~\bibnamefont {Tanuma}}, \bibinfo
  {author} {\bibfnamefont {S.}~\bibnamefont {Fujioka}}, \bibinfo {author}
  {\bibfnamefont {H.}~\bibnamefont {Nishimura}}, \bibinfo {author}
  {\bibfnamefont {K.}~\bibnamefont {Nishihara}}, \bibinfo {author}
  {\bibfnamefont {T.}~\bibnamefont {Kai}}, \bibinfo {author} {\bibfnamefont
  {A.}~\bibnamefont {Sasaki}}, \bibinfo {author} {\bibfnamefont {H.~A.}\
  \bibnamefont {Sakaue}}, \bibinfo {author} {\bibfnamefont {N.}~\bibnamefont
  {Nakamura}},  \emph {et~al.},\ }in\ \href@noop {} {\emph {\bibinfo
  {booktitle} {JPCS}}},\ Vol.\ \bibinfo {volume} {163}\ (\bibinfo
  {organization} {IOP Publishing},\ \bibinfo {year} {2009})\ p.\ \bibinfo
  {pages} {012071}\BibitemShut {NoStop}%
\bibitem [{\citenamefont {D'Arcy}\ \emph
  {et~al.}(2009{\natexlab{a}})\citenamefont {D'Arcy}, \citenamefont {Ohashi},
  \citenamefont {Suda}, \citenamefont {Tanuma}, \citenamefont {Fujioka},
  \citenamefont {Nishimura}, \citenamefont {Nishihara}, \citenamefont {Suzuki},
  \citenamefont {Kato}, \citenamefont {Koike}, \citenamefont {White},\ and\
  \citenamefont {O'Sullivan}}]{DArcy2009transitions}%
  \BibitemOpen
  \bibfield  {author} {\bibinfo {author} {\bibfnamefont {R.}~\bibnamefont
  {D'Arcy}}, \bibinfo {author} {\bibfnamefont {H.}~\bibnamefont {Ohashi}},
  \bibinfo {author} {\bibfnamefont {S.}~\bibnamefont {Suda}}, \bibinfo {author}
  {\bibfnamefont {H.}~\bibnamefont {Tanuma}}, \bibinfo {author} {\bibfnamefont
  {S.}~\bibnamefont {Fujioka}}, \bibinfo {author} {\bibfnamefont
  {H.}~\bibnamefont {Nishimura}}, \bibinfo {author} {\bibfnamefont
  {K.}~\bibnamefont {Nishihara}}, \bibinfo {author} {\bibfnamefont
  {C.}~\bibnamefont {Suzuki}}, \bibinfo {author} {\bibfnamefont
  {T.}~\bibnamefont {Kato}}, \bibinfo {author} {\bibfnamefont {F.}~\bibnamefont
  {Koike}}, \bibinfo {author} {\bibfnamefont {J.}~\bibnamefont {White}}, \ and\
  \bibinfo {author} {\bibfnamefont {G.}~\bibnamefont {O'Sullivan}},\ }\href
  {\doibase 10.1103/PhysRevA.79.042509} {\bibfield  {journal} {\bibinfo
  {journal} {Phys. Rev. A}\ }\textbf {\bibinfo {volume} {79}},\ \bibinfo
  {pages} {042509} (\bibinfo {year} {2009}{\natexlab{a}})}\BibitemShut
  {NoStop}%
\bibitem [{\citenamefont {D'Arcy}\ \emph
  {et~al.}(2009{\natexlab{b}})\citenamefont {D'Arcy}, \citenamefont {Ohashi},
  \citenamefont {Suda}, \citenamefont {Tanuma}, \citenamefont {Fujioka},
  \citenamefont {Nishimura}, \citenamefont {Nishihara}, \citenamefont {Suzuki},
  \citenamefont {Kato}, \citenamefont {Koike}, \citenamefont {O'Connor},\ and\
  \citenamefont {O'Sullivan}}]{DArcy2009identification}%
  \BibitemOpen
  \bibfield  {author} {\bibinfo {author} {\bibfnamefont {R.}~\bibnamefont
  {D'Arcy}}, \bibinfo {author} {\bibfnamefont {H.}~\bibnamefont {Ohashi}},
  \bibinfo {author} {\bibfnamefont {S.}~\bibnamefont {Suda}}, \bibinfo {author}
  {\bibfnamefont {H.}~\bibnamefont {Tanuma}}, \bibinfo {author} {\bibfnamefont
  {S.}~\bibnamefont {Fujioka}}, \bibinfo {author} {\bibfnamefont
  {H.}~\bibnamefont {Nishimura}}, \bibinfo {author} {\bibfnamefont
  {K.}~\bibnamefont {Nishihara}}, \bibinfo {author} {\bibfnamefont
  {C.}~\bibnamefont {Suzuki}}, \bibinfo {author} {\bibfnamefont
  {T.}~\bibnamefont {Kato}}, \bibinfo {author} {\bibfnamefont {F.}~\bibnamefont
  {Koike}}, \bibinfo {author} {\bibfnamefont {A.}~\bibnamefont {O'Connor}}, \
  and\ \bibinfo {author} {\bibfnamefont {G.}~\bibnamefont {O'Sullivan}},\
  }\href {\doibase 10.1088/0953-4075/42/16/165207} {\bibfield  {journal}
  {\bibinfo  {journal} {J. Phys. B}\ }\textbf {\bibinfo {volume} {42}},\
  \bibinfo {pages} {165207} (\bibinfo {year} {2009}{\natexlab{b}})}\BibitemShut
  {NoStop}%
\bibitem [{\citenamefont {Ohashi}\ \emph {et~al.}(2010)\citenamefont {Ohashi},
  \citenamefont {Suda}, \citenamefont {Tanuma}, \citenamefont {Fujioka},
  \citenamefont {Nishimura}, \citenamefont {Sasaki},\ and\ \citenamefont
  {Nishihara}}]{ohashi2010euv}%
  \BibitemOpen
  \bibfield  {author} {\bibinfo {author} {\bibfnamefont {H.}~\bibnamefont
  {Ohashi}}, \bibinfo {author} {\bibfnamefont {S.}~\bibnamefont {Suda}},
  \bibinfo {author} {\bibfnamefont {H.}~\bibnamefont {Tanuma}}, \bibinfo
  {author} {\bibfnamefont {S.}~\bibnamefont {Fujioka}}, \bibinfo {author}
  {\bibfnamefont {H.}~\bibnamefont {Nishimura}}, \bibinfo {author}
  {\bibfnamefont {A.}~\bibnamefont {Sasaki}}, \ and\ \bibinfo {author}
  {\bibfnamefont {K.}~\bibnamefont {Nishihara}},\ }\href@noop {} {\bibfield
  {journal} {\bibinfo  {journal} {J. Phys. B}\ }\textbf {\bibinfo {volume}
  {43}},\ \bibinfo {pages} {065204} (\bibinfo {year} {2010})}\BibitemShut
  {NoStop}%
\bibitem [{\citenamefont {Yatsurugi}\ \emph {et~al.}(2011)\citenamefont
  {Yatsurugi}, \citenamefont {Watanabe}, \citenamefont {Ohashi}, \citenamefont
  {Sakaue},\ and\ \citenamefont {Nakamura}}]{yatsurugi2011euv}%
  \BibitemOpen
  \bibfield  {author} {\bibinfo {author} {\bibfnamefont {J.}~\bibnamefont
  {Yatsurugi}}, \bibinfo {author} {\bibfnamefont {E.}~\bibnamefont {Watanabe}},
  \bibinfo {author} {\bibfnamefont {H.}~\bibnamefont {Ohashi}}, \bibinfo
  {author} {\bibfnamefont {H.~A.}\ \bibnamefont {Sakaue}}, \ and\ \bibinfo
  {author} {\bibfnamefont {N.}~\bibnamefont {Nakamura}},\ }\href@noop {}
  {\bibfield  {journal} {\bibinfo  {journal} {Phys. Scr.}\ }\textbf {\bibinfo
  {volume} {2011}},\ \bibinfo {pages} {014031} (\bibinfo {year}
  {2011})}\BibitemShut {NoStop}%
\bibitem [{\citenamefont {Gu}(2008)}]{gu2008flexible}%
  \BibitemOpen
  \bibfield  {author} {\bibinfo {author} {\bibfnamefont {M.~F.}\ \bibnamefont
  {Gu}},\ }\href@noop {} {\bibfield  {journal} {\bibinfo  {journal} {Can. J.
  Phys.}\ }\textbf {\bibinfo {volume} {86}},\ \bibinfo {pages} {675} (\bibinfo
  {year} {2008})}\BibitemShut {NoStop}%
\bibitem [{\citenamefont {Epp}\ \emph {et~al.}(2010)\citenamefont {Epp},
  \citenamefont {{J. R. Crespo L\'{o}pez-Urrutia}}, \citenamefont {Simon},
  \citenamefont {Baumann}, \citenamefont {Brenner}, \citenamefont {Ginzel},
  \citenamefont {Guerassimova}, \citenamefont {M{\"a}ckel}, \citenamefont
  {Mokler}, \citenamefont {Schmitt} \emph {et~al.}}]{epp2010x}%
  \BibitemOpen
  \bibfield  {author} {\bibinfo {author} {\bibfnamefont {S.~W.}\ \bibnamefont
  {Epp}}, \bibinfo {author} {\bibnamefont {{J. R. Crespo L\'{o}pez-Urrutia}}},
  \bibinfo {author} {\bibfnamefont {M.~C.}\ \bibnamefont {Simon}}, \bibinfo
  {author} {\bibfnamefont {T.}~\bibnamefont {Baumann}}, \bibinfo {author}
  {\bibfnamefont {G.}~\bibnamefont {Brenner}}, \bibinfo {author} {\bibfnamefont
  {R.}~\bibnamefont {Ginzel}}, \bibinfo {author} {\bibfnamefont
  {N.}~\bibnamefont {Guerassimova}}, \bibinfo {author} {\bibfnamefont
  {V.}~\bibnamefont {M{\"a}ckel}}, \bibinfo {author} {\bibfnamefont {P.~H.}\
  \bibnamefont {Mokler}}, \bibinfo {author} {\bibfnamefont {B.~L.}\
  \bibnamefont {Schmitt}},  \emph {et~al.},\ }\href@noop {} {\bibfield
  {journal} {\bibinfo  {journal} {J. Phys. B}\ }\textbf {\bibinfo {volume}
  {43}},\ \bibinfo {pages} {194008} (\bibinfo {year} {2010})}\BibitemShut
  {NoStop}%
\bibitem [{\citenamefont {Cowan}(1981)}]{cowan1981}%
  \BibitemOpen
  \bibfield  {author} {\bibinfo {author} {\bibfnamefont {R.~D.}\ \bibnamefont
  {Cowan}},\ }\href@noop {} {\emph {\bibinfo {title} {{The Theory of Atomic
  Structure and Spectra}}}}\ (\bibinfo  {publisher} {University of California
  Press},\ \bibinfo {year} {1981})\BibitemShut {NoStop}%
\bibitem [{\citenamefont {Lindgren}(1996)}]{lindgren1996relativistic}%
  \BibitemOpen
  \bibfield  {author} {\bibinfo {author} {\bibfnamefont {I.}~\bibnamefont
  {Lindgren}},\ }\href@noop {} {\bibfield  {journal} {\bibinfo  {journal} {Int.
  J. Quantum Chem.}\ }\textbf {\bibinfo {volume} {57}},\ \bibinfo {pages} {683}
  (\bibinfo {year} {1996})}\BibitemShut {NoStop}%
\bibitem [{\citenamefont {Eliav}\ \emph {et~al.}(1995)\citenamefont {Eliav},
  \citenamefont {Kaldor},\ and\ \citenamefont
  {Ishikawa}}]{eliav1995relativistic}%
  \BibitemOpen
  \bibfield  {author} {\bibinfo {author} {\bibfnamefont {E.}~\bibnamefont
  {Eliav}}, \bibinfo {author} {\bibfnamefont {U.}~\bibnamefont {Kaldor}}, \
  and\ \bibinfo {author} {\bibfnamefont {Y.}~\bibnamefont {Ishikawa}},\
  }\href@noop {} {\bibfield  {journal} {\bibinfo  {journal} {Phys. Rev. A}\
  }\textbf {\bibinfo {volume} {51}},\ \bibinfo {pages} {225} (\bibinfo {year}
  {1995})}\BibitemShut {NoStop}%
\bibitem [{\citenamefont {Nandy}\ and\ \citenamefont
  {Sahoo}(2013)}]{nandy2013development}%
  \BibitemOpen
  \bibfield  {author} {\bibinfo {author} {\bibfnamefont {D.~K.}\ \bibnamefont
  {Nandy}}\ and\ \bibinfo {author} {\bibfnamefont {B.~K.}\ \bibnamefont
  {Sahoo}},\ }\href@noop {} {\bibfield  {journal} {\bibinfo  {journal} {Phys.
  Rev. A}\ }\textbf {\bibinfo {volume} {88}},\ \bibinfo {pages} {052512}
  (\bibinfo {year} {2013})}\BibitemShut {NoStop}%
\bibitem [{\citenamefont {Safronova}\ \emph {et~al.}(2014)\citenamefont
  {Safronova}, \citenamefont {Dzuba}, \citenamefont {Flambaum}, \citenamefont
  {Safronova}, \citenamefont {Porsev},\ and\ \citenamefont
  {Kozlov}}]{safronova2014highly}%
  \BibitemOpen
  \bibfield  {author} {\bibinfo {author} {\bibfnamefont {M.~S.}\ \bibnamefont
  {Safronova}}, \bibinfo {author} {\bibfnamefont {V.~A.}\ \bibnamefont
  {Dzuba}}, \bibinfo {author} {\bibfnamefont {V.~V.}\ \bibnamefont {Flambaum}},
  \bibinfo {author} {\bibfnamefont {U.~I.}\ \bibnamefont {Safronova}}, \bibinfo
  {author} {\bibfnamefont {S.~G.}\ \bibnamefont {Porsev}}, \ and\ \bibinfo
  {author} {\bibfnamefont {M.~G.}\ \bibnamefont {Kozlov}},\ }\href@noop {}
  {\bibfield  {journal} {\bibinfo  {journal} {Phys. Rev. Lett.}\ }\textbf
  {\bibinfo {volume} {113}},\ \bibinfo {pages} {030801} (\bibinfo {year}
  {2014})}\BibitemShut {NoStop}%
\bibitem [{\citenamefont {Windberger}\ \emph {et~al.}(2015)\citenamefont
  {Windberger}, \citenamefont {Crespo L\'opez-Urrutia}, \citenamefont {Bekker},
  \citenamefont {Oreshkina}, \citenamefont {Berengut}, \citenamefont {Bock},
  \citenamefont {Borschevsky}, \citenamefont {Dzuba}, \citenamefont {Eliav},
  \citenamefont {Harman}, \citenamefont {Kaldor}, \citenamefont {Kaul},
  \citenamefont {Safronova}, \citenamefont {Flambaum}, \citenamefont {Keitel},
  \citenamefont {Schmidt}, \citenamefont {Ullrich},\ and\ \citenamefont
  {Versolato}}]{windberger2015}%
  \BibitemOpen
  \bibfield  {author} {\bibinfo {author} {\bibfnamefont {A.}~\bibnamefont
  {Windberger}}, \bibinfo {author} {\bibfnamefont {J.~R.}\ \bibnamefont {Crespo
  L\'opez-Urrutia}}, \bibinfo {author} {\bibfnamefont {H.}~\bibnamefont
  {Bekker}}, \bibinfo {author} {\bibfnamefont {N.~S.}\ \bibnamefont
  {Oreshkina}}, \bibinfo {author} {\bibfnamefont {J.~C.}\ \bibnamefont
  {Berengut}}, \bibinfo {author} {\bibfnamefont {V.}~\bibnamefont {Bock}},
  \bibinfo {author} {\bibfnamefont {A.}~\bibnamefont {Borschevsky}}, \bibinfo
  {author} {\bibfnamefont {V.~A.}\ \bibnamefont {Dzuba}}, \bibinfo {author}
  {\bibfnamefont {E.}~\bibnamefont {Eliav}}, \bibinfo {author} {\bibfnamefont
  {Z.}~\bibnamefont {Harman}}, \bibinfo {author} {\bibfnamefont
  {U.}~\bibnamefont {Kaldor}}, \bibinfo {author} {\bibfnamefont
  {S.}~\bibnamefont {Kaul}}, \bibinfo {author} {\bibfnamefont {U.~I.}\
  \bibnamefont {Safronova}}, \bibinfo {author} {\bibfnamefont {V.~V.}\
  \bibnamefont {Flambaum}}, \bibinfo {author} {\bibfnamefont {C.~H.}\
  \bibnamefont {Keitel}}, \bibinfo {author} {\bibfnamefont {P.~O.}\
  \bibnamefont {Schmidt}}, \bibinfo {author} {\bibfnamefont {J.}~\bibnamefont
  {Ullrich}}, \ and\ \bibinfo {author} {\bibfnamefont {O.~O.}\ \bibnamefont
  {Versolato}},\ }\href {\doibase 10.1103/PhysRevLett.114.150801} {\bibfield
  {journal} {\bibinfo  {journal} {Phys. Rev. Lett.}\ }\textbf {\bibinfo
  {volume} {114}},\ \bibinfo {pages} {150801} (\bibinfo {year}
  {2015})}\BibitemShut {NoStop}%
\bibitem [{\citenamefont {Ryabtsev}\ and\ \citenamefont
  {Kononov}(2011)}]{ryabtsev2011resonance}%
  \BibitemOpen
  \bibfield  {author} {\bibinfo {author} {\bibfnamefont {A.~N.}\ \bibnamefont
  {Ryabtsev}}\ and\ \bibinfo {author} {\bibfnamefont {E.~Y.}\ \bibnamefont
  {Kononov}},\ }\href@noop {} {\bibfield  {journal} {\bibinfo  {journal} {Phys.
  Scr.}\ }\textbf {\bibinfo {volume} {84}},\ \bibinfo {pages} {015301}
  (\bibinfo {year} {2011})}\BibitemShut {NoStop}%
\bibitem [{\citenamefont {Penetrante}\ \emph {et~al.}(1991)\citenamefont
  {Penetrante}, \citenamefont {Bardsley}, \citenamefont {DeWitt}, \citenamefont
  {Clark},\ and\ \citenamefont {Schneider}}]{penetrante1991evolution}%
  \BibitemOpen
  \bibfield  {author} {\bibinfo {author} {\bibfnamefont {B.~M.}\ \bibnamefont
  {Penetrante}}, \bibinfo {author} {\bibfnamefont {J.~N.}\ \bibnamefont
  {Bardsley}}, \bibinfo {author} {\bibfnamefont {D.}~\bibnamefont {DeWitt}},
  \bibinfo {author} {\bibfnamefont {M.}~\bibnamefont {Clark}}, \ and\ \bibinfo
  {author} {\bibfnamefont {D.}~\bibnamefont {Schneider}},\ }\href@noop {}
  {\bibfield  {journal} {\bibinfo  {journal} {Phys. Rev. A}\ }\textbf {\bibinfo
  {volume} {43}},\ \bibinfo {pages} {4861} (\bibinfo {year}
  {1991})}\BibitemShut {NoStop}%
\bibitem [{\citenamefont {Brenner}\ \emph {et~al.}(2007)\citenamefont
  {Brenner}, \citenamefont {{J. R. Crespo L\'{o}pez-Urrutia}}, \citenamefont
  {Harman}, \citenamefont {Mokler},\ and\ \citenamefont
  {Ullrich}}]{brenner2007lifetime}%
  \BibitemOpen
  \bibfield  {author} {\bibinfo {author} {\bibfnamefont {G.}~\bibnamefont
  {Brenner}}, \bibinfo {author} {\bibnamefont {{J. R. Crespo
  L\'{o}pez-Urrutia}}}, \bibinfo {author} {\bibfnamefont {Z.}~\bibnamefont
  {Harman}}, \bibinfo {author} {\bibfnamefont {P.~H.}\ \bibnamefont {Mokler}},
  \ and\ \bibinfo {author} {\bibfnamefont {J.}~\bibnamefont {Ullrich}},\
  }\href@noop {} {\bibfield  {journal} {\bibinfo  {journal} {Phys. Rev. A}\
  }\textbf {\bibinfo {volume} {75}},\ \bibinfo {pages} {032504} (\bibinfo
  {year} {2007})}\BibitemShut {NoStop}%
\bibitem [{\citenamefont {Bekker}\ \emph {et~al.}(2015)\citenamefont {Bekker},
  \citenamefont {Versolato}, \citenamefont {Windberger}, \citenamefont
  {Oreshkina}, \citenamefont {Schupp}, \citenamefont {Baumann}, \citenamefont
  {Harman}, \citenamefont {Keitel}, \citenamefont {Schmidt}, \citenamefont
  {Ullrich},\ and\ \citenamefont {{J. R. Crespo
  L\'{o}pez-Urrutia}}}]{Bekker2015a}%
  \BibitemOpen
  \bibfield  {author} {\bibinfo {author} {\bibfnamefont {H.}~\bibnamefont
  {Bekker}}, \bibinfo {author} {\bibfnamefont {O.~O.}\ \bibnamefont
  {Versolato}}, \bibinfo {author} {\bibfnamefont {A.}~\bibnamefont
  {Windberger}}, \bibinfo {author} {\bibfnamefont {N.~S.}\ \bibnamefont
  {Oreshkina}}, \bibinfo {author} {\bibfnamefont {R.}~\bibnamefont {Schupp}},
  \bibinfo {author} {\bibfnamefont {T.~M.}\ \bibnamefont {Baumann}}, \bibinfo
  {author} {\bibfnamefont {Z.}~\bibnamefont {Harman}}, \bibinfo {author}
  {\bibfnamefont {C.~H.}\ \bibnamefont {Keitel}}, \bibinfo {author}
  {\bibfnamefont {P.~O.}\ \bibnamefont {Schmidt}}, \bibinfo {author}
  {\bibfnamefont {J.}~\bibnamefont {Ullrich}}, \ and\ \bibinfo {author}
  {\bibnamefont {{J. R. Crespo L\'{o}pez-Urrutia}}},\ }\href {\doibase
  10.1088/0953-4075/48/14/144018} {\bibfield  {journal} {\bibinfo  {journal}
  {J. Phys. B}\ }\textbf {\bibinfo {volume} {48}},\ \bibinfo {pages} {144018}
  (\bibinfo {year} {2015})}\BibitemShut {NoStop}%
\bibitem [{\citenamefont {Rodrigues}\ \emph {et~al.}(2004)\citenamefont
  {Rodrigues}, \citenamefont {Indelicato}, \citenamefont {Santos},
  \citenamefont {Patt{\'e}},\ and\ \citenamefont
  {Parente}}]{rodrigues2004systematic}%
  \BibitemOpen
  \bibfield  {author} {\bibinfo {author} {\bibfnamefont {G.~C.}\ \bibnamefont
  {Rodrigues}}, \bibinfo {author} {\bibfnamefont {P.}~\bibnamefont
  {Indelicato}}, \bibinfo {author} {\bibfnamefont {J.~P.}\ \bibnamefont
  {Santos}}, \bibinfo {author} {\bibfnamefont {P.}~\bibnamefont {Patt{\'e}}}, \
  and\ \bibinfo {author} {\bibfnamefont {F.}~\bibnamefont {Parente}},\
  }\href@noop {} {\bibfield  {journal} {\bibinfo  {journal} {At. Data. Nucl.
  Data Tables}\ }\textbf {\bibinfo {volume} {86}},\ \bibinfo {pages} {117}
  (\bibinfo {year} {2004})}\BibitemShut {NoStop}%
\bibitem [{\citenamefont {Kramida}\ \emph {et~al.}(2015)\citenamefont
  {Kramida}, \citenamefont {{Yu.~Ralchenko}}, \citenamefont {Reader},\ and\
  \citenamefont {{and NIST ASD Team}}}]{NIST_ASD}%
  \BibitemOpen
  \bibfield  {author} {\bibinfo {author} {\bibfnamefont {A.}~\bibnamefont
  {Kramida}}, \bibinfo {author} {\bibnamefont {{Yu.~Ralchenko}}}, \bibinfo
  {author} {\bibfnamefont {J.}~\bibnamefont {Reader}}, \ and\ \bibinfo {author}
  {\bibnamefont {{and NIST ASD Team}}},\ }\href@noop {} {}\bibinfo
  {howpublished} {{NIST Atomic Spectra Database (ver. 5.3), [Online].
  Available: {\tt{http://physics.nist.gov/asd}} [2016, February 2]. National
  Institute of Standards and Technology, Gaithersburg, MD.}} (\bibinfo {year}
  {2015})\BibitemShut {NoStop}%
\bibitem [{\citenamefont {Sucher}(1980)}]{Suc80}%
  \BibitemOpen
  \bibfield  {author} {\bibinfo {author} {\bibfnamefont {J.}~\bibnamefont
  {Sucher}},\ }\href@noop {} {\bibfield  {journal} {\bibinfo  {journal} {Phys.
  Rev. A}\ }\textbf {\bibinfo {volume} {22}},\ \bibinfo {pages} {348} (\bibinfo
  {year} {1980})}\BibitemShut {NoStop}%
\bibitem [{\citenamefont {Ishikawa}\ \emph {et~al.}(1985)\citenamefont
  {Ishikawa}, \citenamefont {Baretty},\ and\ \citenamefont
  {Binning}}]{IshBarBin85}%
  \BibitemOpen
  \bibfield  {author} {\bibinfo {author} {\bibfnamefont {Y.}~\bibnamefont
  {Ishikawa}}, \bibinfo {author} {\bibfnamefont {R.}~\bibnamefont {Baretty}}, \
  and\ \bibinfo {author} {\bibfnamefont {R.~C.}\ \bibnamefont {Binning}},\
  }\href@noop {} {\bibfield  {journal} {\bibinfo  {journal} {Chem. Phys.
  Lett.}\ }\textbf {\bibinfo {volume} {121}},\ \bibinfo {pages} {130} (\bibinfo
  {year} {1985})}\BibitemShut {NoStop}%
\bibitem [{\citenamefont {Eliav}\ \emph {et~al.}(2005)\citenamefont {Eliav},
  \citenamefont {Vilkas}, \citenamefont {Ishikawa},\ and\ \citenamefont
  {Kaldor}}]{EliVilIsh05}%
  \BibitemOpen
  \bibfield  {author} {\bibinfo {author} {\bibfnamefont {E.}~\bibnamefont
  {Eliav}}, \bibinfo {author} {\bibfnamefont {M.~J.}\ \bibnamefont {Vilkas}},
  \bibinfo {author} {\bibfnamefont {Y.}~\bibnamefont {Ishikawa}}, \ and\
  \bibinfo {author} {\bibfnamefont {U.}~\bibnamefont {Kaldor}},\ }\href@noop {}
  {\bibfield  {journal} {\bibinfo  {journal} {J. Chem. Phys.}\ }\textbf
  {\bibinfo {volume} {122}},\ \bibinfo {pages} {224113} (\bibinfo {year}
  {2005})}\BibitemShut {NoStop}%
\bibitem [{\citenamefont {Malli}\ \emph {et~al.}(1993)\citenamefont {Malli},
  \citenamefont {Da~Silva},\ and\ \citenamefont {Ishikawa}}]{MalSilIsh93}%
  \BibitemOpen
  \bibfield  {author} {\bibinfo {author} {\bibfnamefont {G.~L.}\ \bibnamefont
  {Malli}}, \bibinfo {author} {\bibfnamefont {A.~B.~F.}\ \bibnamefont
  {Da~Silva}}, \ and\ \bibinfo {author} {\bibfnamefont {Y.}~\bibnamefont
  {Ishikawa}},\ }\href@noop {} {\bibfield  {journal} {\bibinfo  {journal}
  {Phys. Rev. A}\ }\textbf {\bibinfo {volume} {47}},\ \bibinfo {pages} {143}
  (\bibinfo {year} {1993})}\BibitemShut {NoStop}%
\bibitem [{\citenamefont {Shabaev}\ \emph {et~al.}(2015)\citenamefont
  {Shabaev}, \citenamefont {Tupitsyn},\ and\ \citenamefont
  {Yerokhin}}]{Shabaev2015_qed}%
  \BibitemOpen
  \bibfield  {author} {\bibinfo {author} {\bibfnamefont {V.~M.}\ \bibnamefont
  {Shabaev}}, \bibinfo {author} {\bibfnamefont {I.~I.}\ \bibnamefont
  {Tupitsyn}}, \ and\ \bibinfo {author} {\bibfnamefont {V.~A.}\ \bibnamefont
  {Yerokhin}},\ }\href {\doibase http://dx.doi.org/10.1016/j.cpc.2014.12.002}
  {\bibfield  {journal} {\bibinfo  {journal} {Comput. Phys. Commun.}\ }\textbf
  {\bibinfo {volume} {189}},\ \bibinfo {pages} {175 } (\bibinfo {year}
  {2015})}\BibitemShut {NoStop}%
\bibitem [{\citenamefont {{J. R. Crespo L\'{o}pez-Urrutia}}\ \emph
  {et~al.}(2002)\citenamefont {{J. R. Crespo L\'{o}pez-Urrutia}}, \citenamefont
  {Beiersdorfer}, \citenamefont {Widmann},\ and\ \citenamefont
  {Decaux}}]{lopez2002visible}%
  \BibitemOpen
  \bibfield  {author} {\bibinfo {author} {\bibnamefont {{J. R. Crespo
  L\'{o}pez-Urrutia}}}, \bibinfo {author} {\bibfnamefont {P.}~\bibnamefont
  {Beiersdorfer}}, \bibinfo {author} {\bibfnamefont {K.}~\bibnamefont
  {Widmann}}, \ and\ \bibinfo {author} {\bibfnamefont {V.}~\bibnamefont
  {Decaux}},\ }\href@noop {} {\bibfield  {journal} {\bibinfo  {journal} {Can.
  J. Phys.}\ }\textbf {\bibinfo {volume} {80}},\ \bibinfo {pages} {1687}
  (\bibinfo {year} {2002})}\BibitemShut {NoStop}%
\bibitem [{\citenamefont {Kobayashi}\ \emph {et~al.}(2015)\citenamefont
  {Kobayashi}, \citenamefont {Kubota}, \citenamefont {Omote}, \citenamefont
  {Komatsu}, \citenamefont {Sakoda}, \citenamefont {Minoshima}, \citenamefont
  {Kato}, \citenamefont {Li}, \citenamefont {Sakaue}, \citenamefont {Murakami}
  \emph {et~al.}}]{kobayashi2015extreme}%
  \BibitemOpen
  \bibfield  {author} {\bibinfo {author} {\bibfnamefont {Y.}~\bibnamefont
  {Kobayashi}}, \bibinfo {author} {\bibfnamefont {K.}~\bibnamefont {Kubota}},
  \bibinfo {author} {\bibfnamefont {K.}~\bibnamefont {Omote}}, \bibinfo
  {author} {\bibfnamefont {A.}~\bibnamefont {Komatsu}}, \bibinfo {author}
  {\bibfnamefont {J.}~\bibnamefont {Sakoda}}, \bibinfo {author} {\bibfnamefont
  {M.}~\bibnamefont {Minoshima}}, \bibinfo {author} {\bibfnamefont
  {D.}~\bibnamefont {Kato}}, \bibinfo {author} {\bibfnamefont {J.}~\bibnamefont
  {Li}}, \bibinfo {author} {\bibfnamefont {H.~A.}\ \bibnamefont {Sakaue}},
  \bibinfo {author} {\bibfnamefont {I.}~\bibnamefont {Murakami}},  \emph
  {et~al.},\ }\href@noop {} {\bibfield  {journal} {\bibinfo  {journal} {Phys.
  Rev. A}\ }\textbf {\bibinfo {volume} {92}},\ \bibinfo {pages} {022510}
  (\bibinfo {year} {2015})}\BibitemShut {NoStop}%
\bibitem [{\citenamefont {Borovik~Jr}\ \emph {et~al.}(2013)\citenamefont
  {Borovik~Jr}, \citenamefont {Gharaibeh}, \citenamefont {Hillenbrand},
  \citenamefont {Schippers},\ and\ \citenamefont {M{\"u}ller}}]{borovik2013}%
  \BibitemOpen
  \bibfield  {author} {\bibinfo {author} {\bibfnamefont {A.}~\bibnamefont
  {Borovik~Jr}}, \bibinfo {author} {\bibfnamefont {M.~F.}\ \bibnamefont
  {Gharaibeh}}, \bibinfo {author} {\bibfnamefont {P.~M.}\ \bibnamefont
  {Hillenbrand}}, \bibinfo {author} {\bibfnamefont {S.}~\bibnamefont
  {Schippers}}, \ and\ \bibinfo {author} {\bibfnamefont {A.}~\bibnamefont
  {M{\"u}ller}},\ }\href@noop {} {\bibfield  {journal} {\bibinfo  {journal} {J.
  Phys. B}\ }\textbf {\bibinfo {volume} {46}},\ \bibinfo {pages} {175201}
  (\bibinfo {year} {2013})}\BibitemShut {NoStop}%
\bibitem [{\citenamefont {Kramida}(2010)}]{kramida2010}%
  \BibitemOpen
  \bibfield  {author} {\bibinfo {author} {\bibfnamefont {A.~E.}\ \bibnamefont
  {Kramida}},\ }\href@noop {} {\bibfield  {journal} {\bibinfo  {journal}
  {Comput. Phys. Commun.}\ }\textbf {\bibinfo {volume} {182}},\ \bibinfo
  {pages} {419} (\bibinfo {year} {2010})}\BibitemShut {NoStop}%
\bibitem [{\citenamefont {Ryabtsev}\ and\ \citenamefont
  {Kononov}(2016)}]{ryabtsev2016eighth}%
  \BibitemOpen
  \bibfield  {author} {\bibinfo {author} {\bibfnamefont {A.~N.}\ \bibnamefont
  {Ryabtsev}}\ and\ \bibinfo {author} {\bibfnamefont {E.~Y.}\ \bibnamefont
  {Kononov}},\ }\href@noop {} {\bibfield  {journal} {\bibinfo  {journal} {Phys.
  Scr.}\ }\textbf {\bibinfo {volume} {91}},\ \bibinfo {pages} {025402}
  (\bibinfo {year} {2016})}\BibitemShut {NoStop}%
\bibitem [{\citenamefont {Bi{\'e}mont}\ \emph {et~al.}(1988)\citenamefont
  {Bi{\'e}mont}, \citenamefont {Cowan},\ and\ \citenamefont
  {Hansen}}]{biemont1988}%
  \BibitemOpen
  \bibfield  {author} {\bibinfo {author} {\bibfnamefont {E.}~\bibnamefont
  {Bi{\'e}mont}}, \bibinfo {author} {\bibfnamefont {R.~D.}\ \bibnamefont
  {Cowan}}, \ and\ \bibinfo {author} {\bibfnamefont {J.~E.}\ \bibnamefont
  {Hansen}},\ }\href@noop {} {\bibfield  {journal} {\bibinfo  {journal} {Phys.
  Scr.}\ }\textbf {\bibinfo {volume} {37}},\ \bibinfo {pages} {850} (\bibinfo
  {year} {1988})}\BibitemShut {NoStop}%
\end{thebibliography}%

%\newpage
%\section*{DISCUSSION MATERIAL}
%
%\begin{figure*}
%\includegraphics[width=18cm]{images/Sn_chargestateladder.pdf}
%\caption{Projection of collections of bright lines onto the voltage ``y'' axis of charge states identified through simultaneous EUV measurements.}
%\end{figure*}%
%
%
%\begin{figure*}
%\includegraphics{images/euv_optical.pdf}
%\caption{Composite spectral map of Sn ions interpolated between discrete voltage steps (see main text) showing optical and EUV data. The orange lines represent spectra taken at the maximum of the fluorescence yield of a specific charge state.}
%\end{figure*}%
%
%\begin{figure*}
%\includegraphics{images/optical_ladder.pdf}
%\caption{Composite spectral map of Sn ions interpolated between discrete voltage steps (see main text) with projection of collections of bright lines onto the voltage ``y'' axis. The orange lines represent spectra taken at the maximum of the fluorescence yield of a specific charge state.}
%\end{figure*}%

\end{document}